\documentclass{article}

\PassOptionsToPackage{numbers,sort&compress}{natbib}
\usepackage[main, final]{neurips_2026}

\usepackage[utf8]{inputenc}
\usepackage[T1]{fontenc}
\usepackage{hyperref}
\usepackage{xurl}
\usepackage{url}
\usepackage{booktabs}
\usepackage{amsfonts}
\usepackage{amsmath}
\usepackage{amssymb}
\usepackage{mathtools}
\usepackage{amsthm}
\usepackage{nicefrac}
\usepackage{microtype}
\usepackage{xcolor}
\usepackage{graphicx}
\usepackage{subcaption}
\usepackage{soul}
\usepackage{xspace}
\usepackage{multirow}
\usepackage{listings}
\usepackage{wrapfig}
\usepackage{pifont}

\usepackage{tabularx}

\definecolor{mydarkblue}{rgb}{0,0.08,0.45}
\hypersetup{
  pdfsubject={Submission to the Conference on Neural Information Processing Systems 2026},
  pdfborder=0 0 0,
  pdfpagemode=UseNone,
  colorlinks=true,
  linkcolor=mydarkblue,
  citecolor=mydarkblue,
  filecolor=mydarkblue,
  urlcolor=mydarkblue
}

\newcommand{\sys}{HetCCL\xspace}
\definecolor{codegray}{rgb}{0.5,0.5,0.5}
\lstdefinestyle{p2p}{
  language=C++,
  basicstyle=\ttfamily\footnotesize,
  numbers=left,
  numberstyle=\scriptsize\color{codegray},
  numbersep=5pt,
  xleftmargin=10pt,
  emph={cudaMalloc, hipMalloc},
  emphstyle=\color{red},
}
\lstdefinestyle{tacc}{
  language=C++,
  basicstyle=\ttfamily\footnotesize,
  numbers=left,
  numberstyle=\scriptsize\color{codegray},
  keywordstyle=\color{blue}\bfseries,
  morekeywords={function, if, else},
}

\usepackage[capitalize,noabbrev]{cleveref}

\theoremstyle{plain}

\theoremstyle{definition}

\theoremstyle{remark}

\title{\sys: Efficient LLM Training \\ on Heterogeneous Vendor GPUs}

\author{%
\textbf{Heehoon Kim}\textsuperscript{1}\thanks{Equal contribution.
\textsuperscript{\ensuremath{\dagger}}This work was done while a Ph.D. student at Seoul National University.}\,
\textsuperscript{\ensuremath{\dagger}}
\quad \textbf{Chris Jaehwan Lee}\textsuperscript{2}\footnotemark[1]
\quad \textbf{Taejeoung Kim}\textsuperscript{3}
\quad \textbf{Jongwon Park}\textsuperscript{2}\\
\textbf{Jinpyo Kim}\textsuperscript{2}
\quad \textbf{Pyongwon Suh}\textsuperscript{3}
\quad \textbf{Ryan H. Choi}\textsuperscript{3}
\quad \textbf{Sangwoo Lee}\textsuperscript{3}
\quad \textbf{Jaejin Lee}\textsuperscript{2,4}\\[0.5em]
\small\textsuperscript{1}Moreh Inc.\quad
\small\textsuperscript{2}Dept. of Computer Science and Engineering, Seoul National University\\
\small\textsuperscript{3}Samsung Research\quad
\small\textsuperscript{4}Graduate School of Data Science, Seoul National University\\
\small\texttt{heehoon.kim@moreh.io, \{jaehwan,jongwon.park,jinpyo\}@aces.snu.ac.kr}\\
\small\texttt{\{tj76.kim,pyongwon.suh,ryan.choi,swa.lee\}@samsung.com, jaejin@snu.ac.kr}\\
}

\begin{document}

\maketitle

\begin{abstract}
The rapid growth of large language models is driving organizations to expand their GPU clusters, often with GPUs from multiple vendors. However, current deep learning frameworks lack support for communication across heterogeneous vendor GPUs, leading to inefficiency and higher costs. We present \sys, a collective communication library that unifies vendor-specific backends and enables RDMA-based communication across GPUs without requiring driver modifications. \sys introduces two novel mechanisms that enable cross-vendor communication while leveraging optimized vendor libraries, NVIDIA NCCL and AMD RCCL. Evaluations on a multi-vendor GPU cluster show that \sys matches NCCL and RCCL performance in homogeneous setups while extending support to heterogeneous environments, enabling practical and efficient training across NVIDIA and AMD GPUs without changes to existing deep learning applications.
\end{abstract}
\section{Introduction}
The recent emergence of trillion-scale deep learning (DL) models~\cite{fedus2022switch, achiam2023gpt, du2022glam} has enabled solutions to previously intractable problems. A key enabler behind this advancement is the high computational capability of heterogeneous cluster systems equipped with various hardware accelerators (e.g., GPUs, NPUs, and FPGAs), which makes it feasible to train and run inference on such massive models. Among these systems, GPU-based platforms play a dominant role in DL, with NVIDIA and AMD GPUs serving as the de facto standard~\cite{otterness2020amd, lehdonvirta2025measuring}.

To enable efficient training and inference of DL models across multiple GPUs, a variety of parallelization techniques have been developed. The most representative approaches are data parallelism~\cite{rajbhandari2020zero} 
and model parallelism, including tensor, pipeline, and expert parallelism~\cite{shoeybi2019megatron, huang2019gpipe, shazeer2017outrageously}. Although these techniques differ in how computations and data are distributed across GPUs, they share a common requirement for efficient coordination across devices. This requirement has become increasingly important as modern ML systems rely on parallelization to improve hardware utilization~\cite{narayanan2021efficient, singh2023hybrid}.


In multi-GPU parallel training or inference, inter-GPU communication is crucial for sharing computation results and synchronizing model states. Collective communication operations, such as All-Reduce, All-Gather, and Reduce-Scatter, are commonly employed for this purpose. Consequently, most accelerator vendors provide optimized collective communication libraries (CCLs) tailored to their hardware. Examples include NVIDIA NCCL~\cite{nvidia_nccl}, AMD RCCL~\cite{amd_rccl}, and Intel oneCCL~\cite{intel_oneccl}.

Training large-scale models typically requires GPU clusters with sufficient compute capacity. In practice, many academic and research institutions expand their clusters by acquiring GPU systems in a piecemeal fashion~\cite{park2020hetpipe, ding2021hetseq}. This incremental procurement approach allows continued use of existing resources while integrating newer hardware. As a result, the infrastructure becomes heterogeneous, comprising various types of GPUs, sometimes even from different vendors. 

While GPUs from different vendors share similar design philosophies and architectures, they require distinct frameworks and toolchains, making cross-vendor software development challenging. Thanks to ongoing efforts across the DL ecosystem, full-featured DL frameworks are now available for each vendor's GPUs. For instance, PyTorch~\cite{paszke2019pytorch} can be configured to support either NVIDIA or AMD GPUs. However, parallel training across GPUs from different vendors remains infeasible because of the incompatibility of their communication backends. DL frameworks depend on vendor-specific libraries, such as NCCL for NVIDIA GPUs and RCCL for AMD GPUs, which are not interoperable.

As DL workloads scale to larger sizes, the ability to achieve fast and efficient communication across heterogeneous GPUs is becoming a crucial factor for building practical and cost-effective AI infrastructure. Investigating such communication, particularly between NVIDIA and AMD GPUs, is not only timely but also essential for the advancement of the next generation of distributed ML systems. For this purpose, this paper presents \sys, a novel CCL that supports GPUs from multiple vendors. To the best of our knowledge, \sys is the first work to enable transparent utilization of all GPUs in a multi-vendor heterogeneous cluster. No source code modifications are required at any level, including the driver, runtime, compiler, and application. In particular, it supports NVIDIA and AMD GPUs, the two leading vendors by market share, which together dominate the accelerator market 
(approximately 88\% and 12\%, respectively)~\cite{lehdonvirta2025measuring}. By replacing original communication backends (i.e., NCCL and RCCL) with \sys, existing distributed training code written in DL frameworks (e.g., PyTorch) can use both vendors' GPUs without further modifications.

The key contributions are summarized as follows:

\begin{itemize}
\item We propose \sys, the first cross-vendor CCL that enables DL model training on heterogeneous clusters with both NVIDIA and AMD GPUs.

\item We show that direct RDMA communication between GPUs from different vendors is possible without modifying GPU drivers or runtimes.

\item We describe the design and implementation of efficient collective communication between GPUs from different vendors, with an emphasis on challenges, such as abstracting heterogeneous runtime APIs and integrating vendor-specific code into a unified framework.

\item We evaluate the training performance of large language models on a multi-vendor GPU cluster and demonstrate that \sys achieves higher aggregate throughput than homogeneous baselines while mitigating straggler effects and preserving model accuracy.
\end{itemize}

We discuss the limitations of this work and outline directions for future research in Appendix~\ref{sec:limitations}.
\begin{figure}[t]
\centering
\begin{subfigure}{0.49\linewidth}
  \centering
  \includegraphics[width=0.85\linewidth]{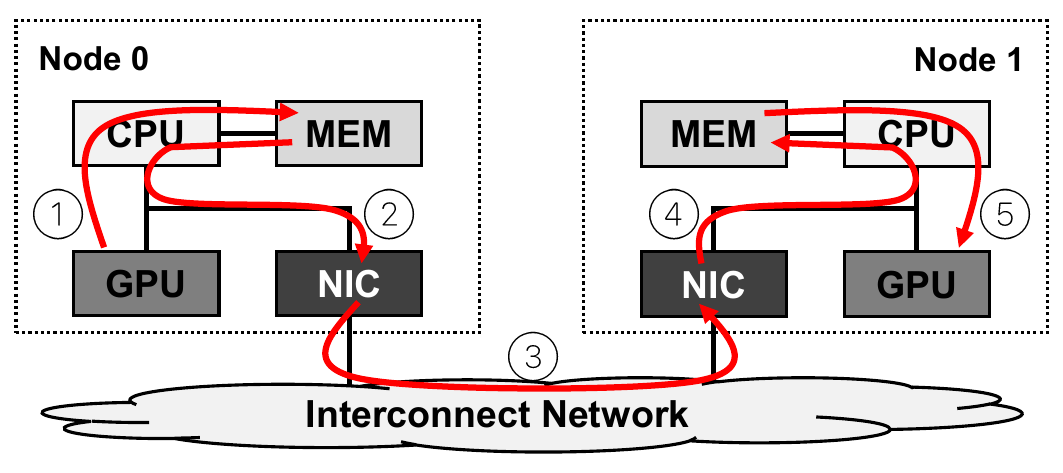}
  \vspace{-0.3\baselineskip}
  \caption{Without RDMA}
  \label{fig:comm-across-nodes-a}
\end{subfigure}
\hfill
\begin{subfigure}{0.49\linewidth}
  \centering
  \includegraphics[width=0.85\linewidth]{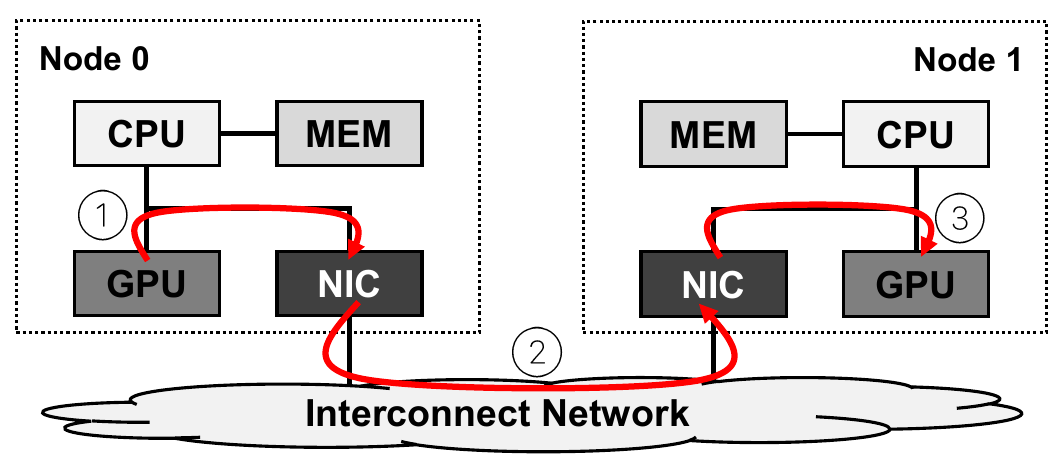}
  \vspace{-0.3\baselineskip}
  \caption{With RDMA}
  \label{fig:comm-across-nodes-b}
\end{subfigure}

\vspace{-0.5ex}
\caption{Communication between GPUs across nodes.}
\label{fig:comm-across-nodes}
\end{figure}

\section{Background and Related Work}
This section provides an overview of GPU communication mechanisms and reviews relevant prior work. An extended discussion is provided in Appendix~\ref{sec:extended-related-work}.

\subsection{Communication Mechanisms of GPUs}

\paragraph{Intra-node communication.}
GPUs within the same node communicate via either \textit{peer-to-peer (P2P)} or \textit{shared memory (SHM)} mode, depending on their interconnect and PCIe topology. While P2P enables direct GPU-to-GPU transfers and offers the highest performance, it may be unavailable or inefficient across PCIe root complexes, in which case communication falls back to SHM, which stages data through host memory and provides lower throughput~\cite{hu2025demystifying}.

\paragraph{Inter-node communication and RDMA.}
In a multi-node environment, GPUs located on different nodes cannot directly access each other's memory. The most basic inter-node transfer mechanism relies on the host memory as a staging area at both the sending and receiving ends (Figure~\ref{fig:comm-across-nodes-a}).
This approach incurs memory copy overhead and suffers from the limited bandwidth of the host memory. 
For faster inter-node communication between GPUs, modern GPUs support \textit{Remote Direct Memory Access} (RDMA)~\cite{recio2007remote}, which allows a network interface card (NIC), such as InfiniBand (IB)~\cite{pfister2001introduction} and RoCE~\cite{kaur2013rdma}, to access GPU memory directly, bypassing the CPU (Figure~\ref{fig:comm-across-nodes-b}). The capability is often marketed under vendor-specific names, such as NVIDIA's GPUDirect~\cite{nvidia_gpudirect} and AMD's DirectGMA~\cite{amd_directgma}. Enabling RDMA requires that the device memory be allocated using vendor-specific APIs (e.g., \texttt{(cuda/hip)Malloc}) and be registered with the IB Verbs API~\cite{macarthur2017integrated}. The registration enables RDMA-capable NICs to directly access GPU memory regions via Verbs' \texttt{send/recv} API.


\subsection{Existing CCLs for GPUs}

\paragraph{Vendor-provided libraries.}
NCCL~\cite{nvidia_nccl} is the most widely used CCL optimized for NVIDIA GPU environments. RCCL~\cite{amd_rccl}, developed by AMD, is a HIP-based implementation that mirrors much of NCCL’s design and interface, but is optimized for AMD GPU systems. Intel’s oneCCL~\cite{intel_oneccl} targets heterogeneous CPU-GPU systems, but it is restricted to Intel hardware. MSCCL++~\cite{shah2025msccl++, microsoft_mscclpp}, proposed by Microsoft, is a programmable framework that can be compiled for either NVIDIA or AMD GPUs, but does not support communication across GPUs from different vendors. Meta developed TorchComms~\cite{si2025collective, meta_torchcomms} to improve the scalability of collective communication within large-scale NVIDIA GPU clusters. These vendor libraries are limited to homogeneous GPU environments and do not allow communication between GPUs from different vendors.

\paragraph{Other libraries.}
MPI~\cite{walker1996mpi} is a widely used communication standard in HPC, but is not specifically optimized for the collective communication patterns of modern accelerator-based DL workloads. TCCL~\cite{kim2024tccl} is an extension of NCCL that optimizes communication for PCIe-dependent systems by profiling and selecting low-congestion paths; however, it is limited to NVIDIA GPUs. HiCCL~\cite{hidayetoglu2025hiccl} builds hierarchical collectives from primitive operations with hierarchy-aware optimizations. UCCL~\cite{zhou2025extensible} focuses on an extensible transport layer that decouples collective libraries from underlying network mechanisms to improve portability and flexibility. Nonetheless, neither HiCCL nor UCCL supports executing a single collective operation across GPUs from different vendors.

\begin{figure}[t]
\centering
\begin{subfigure}{0.532\linewidth}
\centering
  \includegraphics[width=\linewidth]{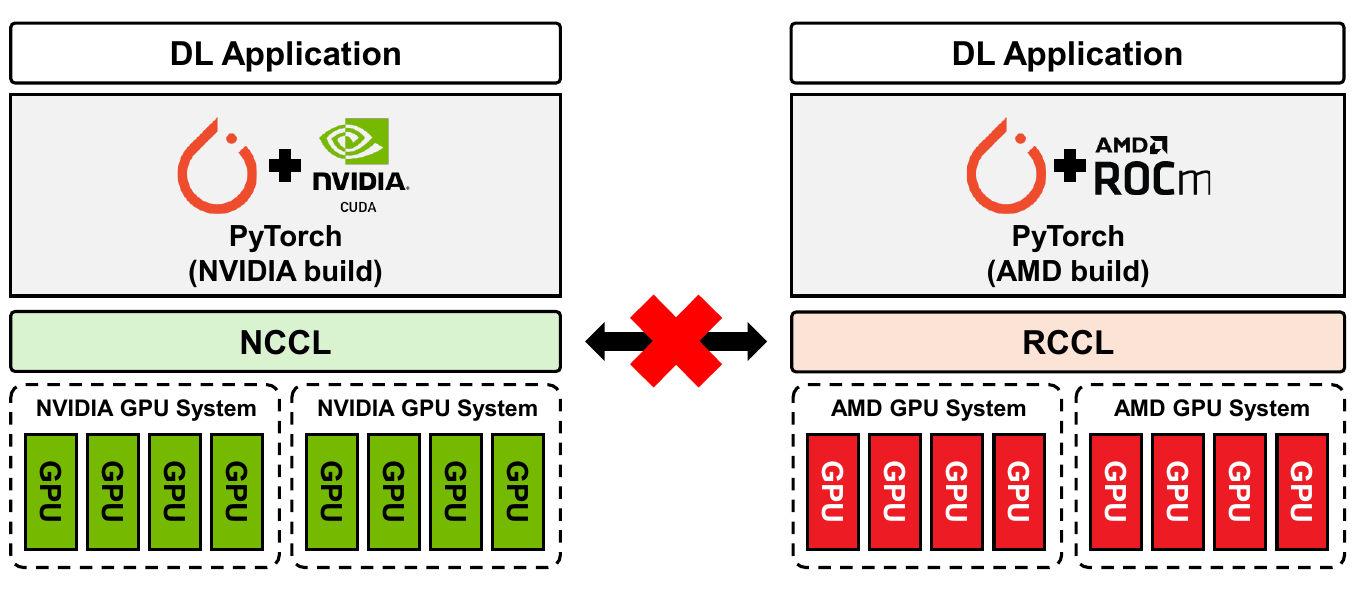}
  \vspace{-1.3\baselineskip}
  \caption{Previous work}
  \label{fig:pytorch-without-system}
\end{subfigure}
\hfill
\begin{subfigure}{0.454\linewidth}
\centering
  \includegraphics[width=\linewidth]{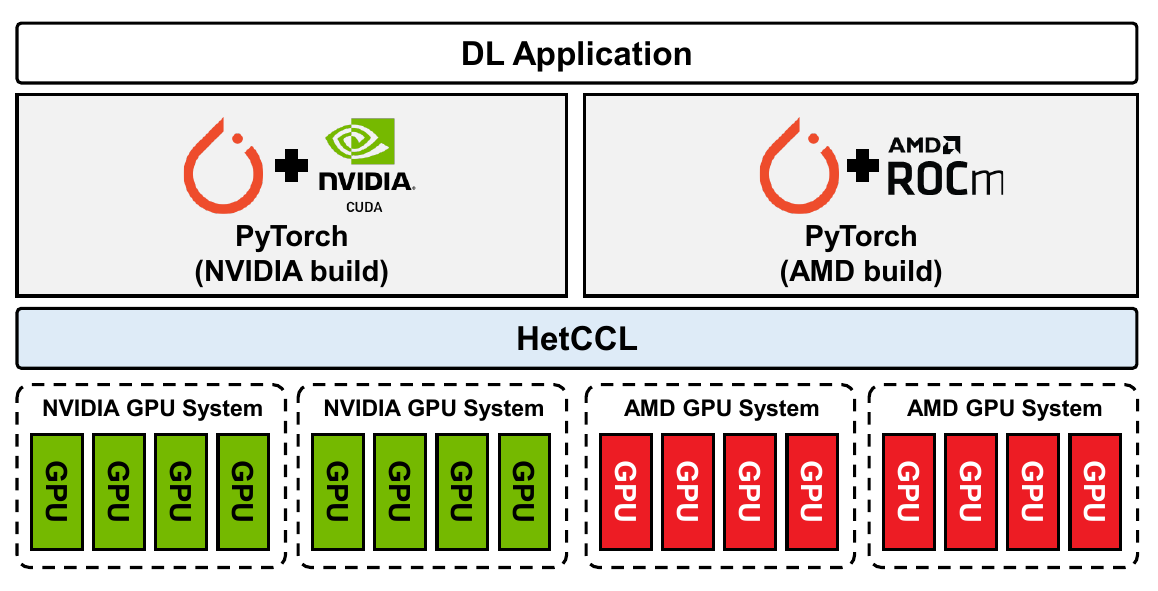}
  \vspace{-1.3\baselineskip}
  \caption{This work (\sys)}
  \label{fig:pytorch-with-system}
\end{subfigure}

\vspace{-0.5ex}
\caption{Comparison between the previous approach and \sys.}
\label{fig:system-overview}
\end{figure}

\section{Motivation}

Efficient coordination among heterogeneous GPUs has become increasingly important as large language model (LLM) training compute scales and clusters are expanded to integrate new hardware while retaining existing GPUs. This process results in heterogeneous deployments across GPU generations or capabilities~\cite{park2020hetpipe, li2022amp}. Unlike hyperscalers that can provision massive homogeneous GPU clusters, academic and institutional mid-scale clusters often rely on incrementally expanded mixed-resource clusters, motivating studies on heterogeneous GPU systems~\cite{li2022amp, um2024metis, ding2021hetseq, yi2020optimizing, park2020hetpipe} that optimize parallelization and load balancing across diverse accelerators as described in Appendix~\ref{subsec:dl_on_hetero_gpus}.

Recently, heterogeneous deployments across GPU vendors are also emerging. Cost and supply constraints of dominant vendors have accelerated the adoption of alternative GPUs from other manufacturers~\cite{narayanan2020heterogeneity}. This trend is evident in real-world infrastructure: major cloud providers~\cite{aws_amd, azure_amd, oracle_amd} and cluster orchestration platforms~\cite{kubernetes_gpu, openshift_gpu} now offer both NVIDIA and AMD GPUs.

However, prior work on heterogeneous GPU systems primarily targets single-vendor heterogeneity (as shown in Table~\ref{tab:hetero_training_scale}), such as different generations of GPUs from the same vendor, and distributed training across multi-vendor GPUs remains largely unsupported. As illustrated in Figure~\ref{fig:pytorch-without-system}, DL frameworks like PyTorch rely on vendor-specific backends (e.g., NCCL for NVIDIA and RCCL for AMD), which prevents interoperability between heterogeneous GPUs within a single training job. Thus, the absence of a cross-vendor CCL remains a fundamental barrier to building flexible, cost-effective ML systems. To overcome this limitation, we propose \sys, which unifies vendor-specific stacks and enables a single training system to run across NVIDIA and AMD GPUs as illustrated in Figure~\ref{fig:pytorch-with-system}.
\section{The \sys Design}

In this section, we describe the design of \sys.
We first present four intertwined challenges in designing a cross-vendor CCL, followed by the key insights underlying \sys.
We then detail \sys{}'s approach, which is implemented through two novel mechanisms.
Finally, we discuss how \sys enables efficient training without code modification or load imbalance.

\subsection{Challenges}

\paragraph{Near-native performance in homogeneous settings.}
Maintaining near-native performance in homogeneous environments is a strict requirement that necessitates the use of vendor-provided CCLs. These libraries are closely aligned with GPU microarchitectures and include extensive vendor-specific optimizations. Although recent MPI implementations~\cite{gabriel2004open, panda2021mvapich} provide broad portability, they typically do not achieve the performance levels of NCCL or RCCL for large-scale data transfers~\cite{chen2023mpi}. Similarly, most previous CCLs~\cite{kim2024tccl, si2025collective} are built on vendor-provided libraries rather than serving as complete replacements. This challenge is intensified by the close integration of NCCL and RCCL within most DL frameworks~\cite{paszke2019pytorch, rasley2020deepspeed, shoeybi2019megatron} and by the rapid advancement of vendor libraries, which frequently introduce new hardware features and optimizations~\cite{hu2025demystifying}. Consequently, although possible, developing a new library from the ground up would be inefficient and inherently limited in performance.

\paragraph{High performance in heterogeneous settings.}
Achieving performance comparable to homogeneous communication in heterogeneous environments remains challenging without high-performance transfer mechanisms. Near-homogeneous performance is typically achievable only through zero-copy RDMA. In contrast, host-mediated relay approaches, such as those implemented in KAITIAN~\cite{lin2025kaitian}, lead to substantial performance degradation. Moreover, RDMA provides only data movement primitives and does not address the scheduling or synchronization of collective operations. As a result, critical coordination challenges persist in heterogeneous collectives.

\paragraph{Heterogeneity in software stacks across vendors.}
Cross-vendor GPU communication is further complicated by fundamental discrepancies in software stacks, including differences in drivers, runtime APIs, and compilation toolchains. Although libraries such as MSCCL++~\cite{microsoft_mscclpp} and TorchComms~\cite{meta_torchcomms} claim to support both NVIDIA and AMD GPUs, they rely on compile-time selection of a single backend (NCCL or RCCL), which prevents simultaneous cross-vendor execution. This limitation arises because CUDA and HIP expose distinct runtime interfaces, leading these libraries to include only one vendor-specific code path in the generated binary (i.e., host code). Additionally, GPU kernels must be compiled with vendor-specific compilers (e.g., \texttt{nvcc} or \texttt{hipcc}), producing assembly code compatible only with the corresponding hardware. MSCCL++ and TorchComms employ a single compiler toolchain at build time. As a result, existing designs inherently restrict execution to a single vendor, and separate vendor-specific builds would therefore need a shared control protocol to form a cross-vendor collective.

\paragraph{Performance disparity and stragglers in heterogeneous collectives.}
Even when heterogeneous communication achieves performance close to that of a homogeneous setup, collective operations remain constrained by the slowest device in the group. This performance disparity leads to straggler effects, intensified by the frequent collective communication operations required in distributed training. It can substantially reduce throughput in large-scale DL workloads such as LLMs.

\subsection{Observation}


\begin{wrapfigure}{r}{0.50\linewidth}
    \centering
    \vspace{-1em}
    \includegraphics[width=\linewidth]{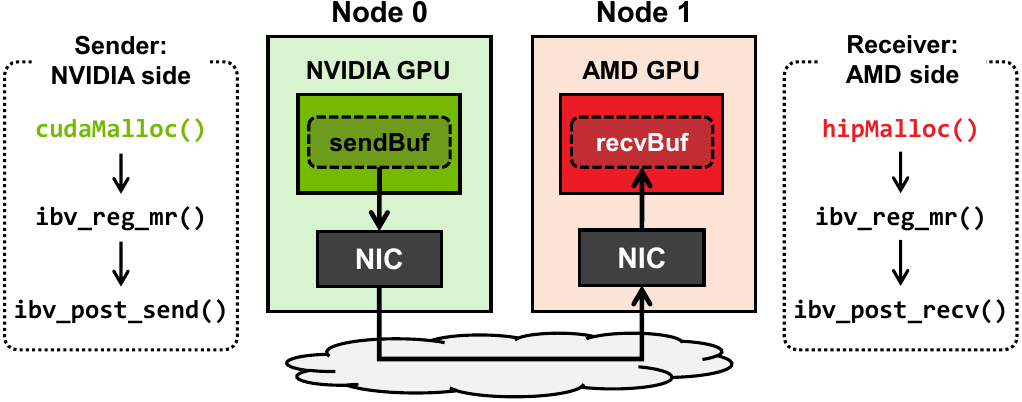}
    \caption{RDMA-based communication between NVIDIA and AMD GPUs in different nodes.}
    \label{fig:hetero-p2p}
    \vspace{-1em}
\end{wrapfigure}

There is a common misconception that RDMA between GPUs from different vendors requires special modifications, such as reimplementing GPU drivers.
However, we observe that once a memory region—whether in host memory or device memory managed by different GPU architectures—is properly registered with a NIC, the underlying RDMA stack abstracts away vendor-specific differences.
We also find that device memory is interoperable across vendors without requiring any data layout transformation (e.g., endianness conversion).
As a result, heterogeneous inter-GPU RDMA is feasible without modifying GPU drivers or runtimes, provided that GPU memory is properly registered with RDMA-capable NICs. Once GPU memory is exposed to the PCIe address space via GPUDirect or DirectGMA, data transfers are handled by the RDMA stack in a vendor-agnostic manner, and the receiver does not need to know the sender’s GPU vendor or memory type.
Figure~\ref{fig:hetero-p2p} illustrates this mechanism with an example where one node is equipped with an NVIDIA GPU and the other with an AMD GPU.
Each GPU allocates communication buffers using its own runtime API (i.e., \texttt{(cuda/hip)Malloc}), registers them with the local NIC via \texttt{ibv\_reg\_mr}, and performs RDMA \texttt{send/recv} operations.
Here, standard RDMA setup procedures (e.g., QP creation and completion polling) are omitted for brevity. With modern hardware and driver support, this mechanism can be applied broadly across heterogeneous systems without requiring cross-vendor driver integration. When RDMA is unavailable, the system can gracefully fall back to staging the data through the host memory. 

\paragraph{Goals.}
\sys aims to enable efficient collective communication across cross-vendor GPUs, pursuing three goals:
\begin{itemize}
    \item In homogeneous settings, \sys must preserve near-native intra-node communication performance comparable to vendor-provided CCLs.
    \item In heterogeneous settings, \sys should achieve performance close to that of homogeneous systems by leveraging RDMA-based inter-node communication.
    \item \sys must maintain backward compatibility with existing DL frameworks, enabling drop-in integration without requiring any code modifications.
\end{itemize}

\paragraph{Key Approach.}
The key insight behind \sys is to decouple cross-vendor collective coordination from vendor-specific execution.
Instead of reimplementing a new library, \sys adapts NCCL and RCCL collective algorithms and protocols into a common host-side implementation. Each rank loads kernels compiled for its local GPU vendor, allowing ranks from both vendors to participate in one collective without invoking NCCL or RCCL as separate libraries, thereby preserving vendor-native optimizations, remaining robust to future updates, and enabling drop-in integration with existing DL frameworks.
To realize this design, \sys employs (1) a runtime API abstraction layer across vendors and (2) a platform-specific device code compilation strategy for multi-vendor kernel execution.
This design further leverages our observation that RDMA enables efficient data movement across heterogeneous GPUs, allowing \sys to achieve near-homogeneous performance in inter-node settings.
In addition, \sys mitigates performance imbalance across heterogeneous GPUs through GPU-aware micro-batch size adjustment, reducing straggler effects during collective synchronization.

    


\subsection{Multi-Vendor Runtime API Abstraction}

\sys leverages code from both NCCL and RCCL to facilitate high-performance communication across GPUs from different vendors. Although these libraries are developed separately, they share a similar structure and functionality. For example, their analogous high-level structures—specifically, the host-side logic for partitioning and communicating data—facilitate their integration, allowing \sys to retain their collective algorithms and protocols in a common host-side control plane. Specifically, \sys uses the existing ring and tree algorithms, with each GPU performing reductions using its vendor-specific kernel. Intra-node P2P or SHM hops and inter-node RDMA hops proceed in the same pipeline, with network hops progressed by existing proxy threads. However, the primary challenge is that each library depends on a distinct set of runtime APIs and device kernels. This limits existing CCLs that support multiple vendors, such as MSCCL++ and TorchComms, which rely on compile-time macros to select a single vendor-specific API, thereby including only one of CUDA or HIP in the host-side library build and preventing simultaneous cross-vendor execution. \sys tackles these differences through a series of unification mechanisms.


\begin{figure}[t]
    \centering
    \begin{minipage}[t]{0.43\linewidth}
        \centering
        \includegraphics[width=1.0\linewidth]{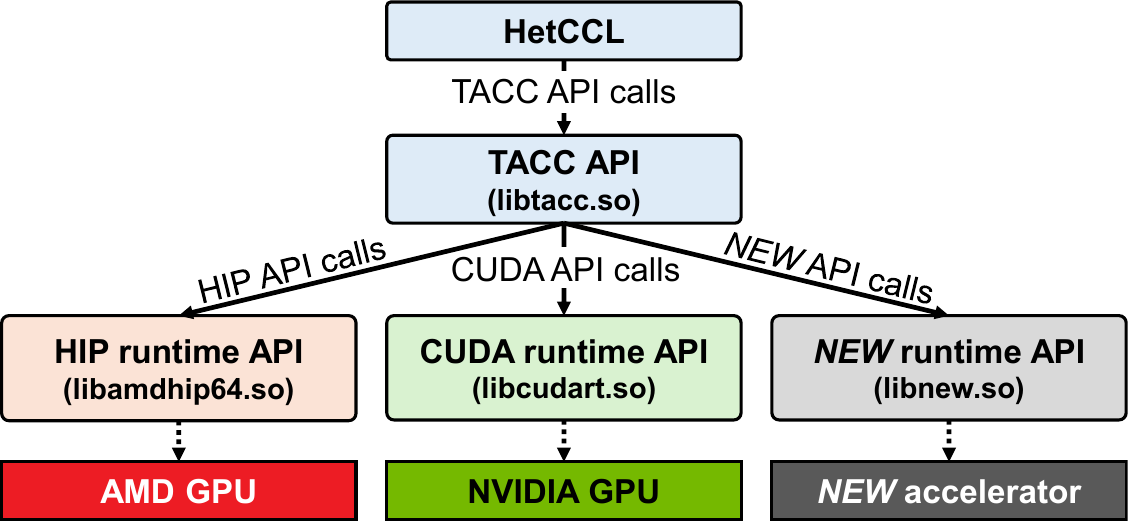}
        \caption{Runtime API abstraction layer.}
        \label{fig:runtime-api-abstraction}
    \end{minipage}
    \hfill
    \begin{minipage}[t]{0.55\linewidth}
        \centering
        \includegraphics[width=1.0\linewidth]{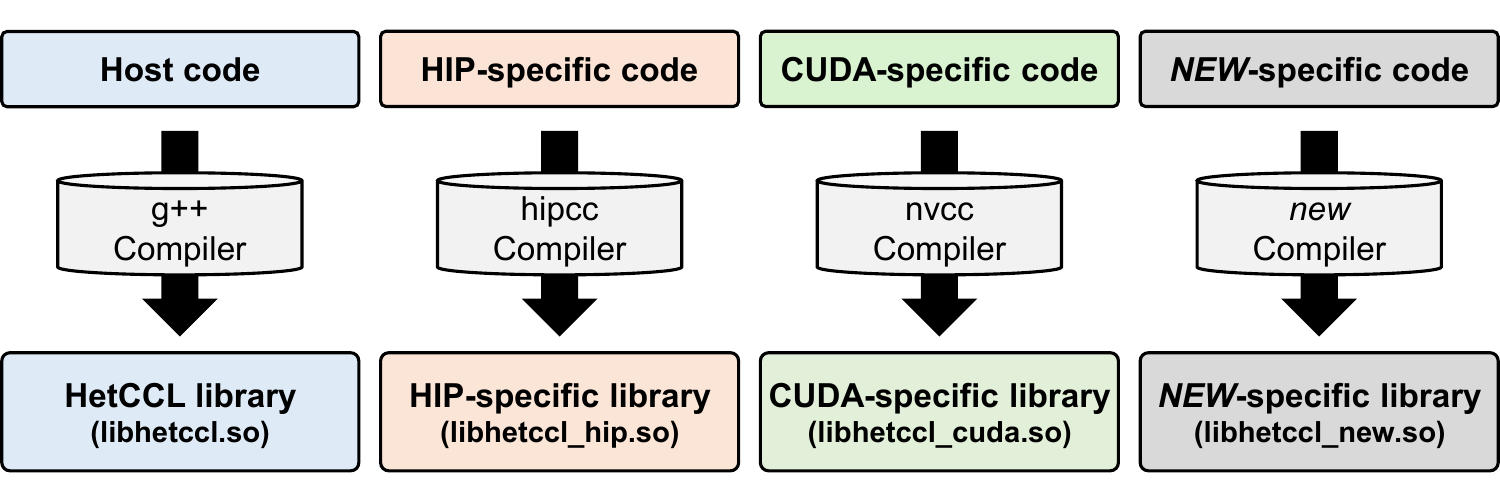}
        \caption{Device code compilation mechanism.}
        \label{fig:device_code_compilation}
    \end{minipage}
\end{figure}

\sys introduces a runtime API abstraction layer called \textit{TACC}. TACC unifies the CUDA and HIP runtime APIs under a common platform-independent interface, as shown in Figure~\ref{fig:runtime-api-abstraction}. Instead of directly calling platform-specific API functions, \sys invokes TACC APIs such as \texttt{taccMalloc} and \texttt{taccMemcpy}. During initialization, TACC automatically detects the platform through \texttt{taccSetPlatformAuto} (or selects it based on user input) and configures a platform-specific function table. Subsequent TACC API calls are then forwarded to the corresponding CUDA or HIP runtime APIs (e.g.,  \texttt{cudaMalloc} or \texttt{hipMalloc}). We list the TACC APIs in Appendix~\ref{sec:tacc-api-functions}.

TACC covers the CUDA and HIP runtime APIs used by NCCL and RCCL, allowing \sys to support multiple platforms without modifying the core communication logic. This design also simplifies future extension to other accelerators, since platform-specific changes can be localized within the abstraction layer. Potential feature gaps between NCCL and RCCL are not a fundamental limitation, because RCCL continuously incorporates new NCCL features through its forked codebase~\cite{amd_rccl}, while \sys relies on native NCCL or RCCL for vendor-local collectives when necessary.

\begin{figure}[ht] 
    \begin{minipage}{\linewidth}
    \centering
    \includegraphics[width=0.84\linewidth]{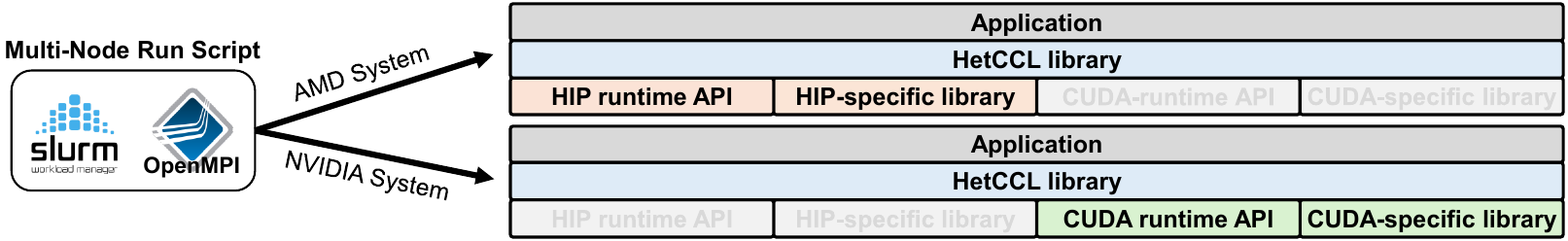}
    \caption{Multi-node execution flow on heterogeneous vendor GPU clusters.}
    \label{fig:launch_system}
    \end{minipage}
\end{figure}

\subsection{Platform-Specific Device Code Compilation}

Unlike runtime API functions, device code (or GPU kernels) cannot be abstracted because it requires compilation with vendor compilers, \texttt{nvcc} for NVIDIA and \texttt{hipcc} for AMD GPUs. Unlike prior work (e.g., MSCCL++ and TorchComms), which requires single-vendor compilation, \sys enables multi-vendor compilation by decoupling device-dependent code into standalone shared libraries independently compiled for each GPU platform.
Figure~\ref{fig:device_code_compilation} illustrates the compilation workflow of \sys. The host-side implementation is shared, while the CUDA and HIP device libraries are compiled separately. At execution time, \sys dynamically loads the appropriate backend library, resolves its kernel entry points, and issues kernel launches through a function pointer table initialized once at startup. Each abstracted kernel launch resolves to a single indirect call, incurring negligible overhead compared with microsecond-to-millisecond collective operations. This approach decouples platform-specific compilation from the core communication logic, allowing seamless extension to future accelerators. Only the corresponding kernel library needs to be implemented and loaded for new platforms.

\subsection{Zero Code Modification}

To avoid modifying application code, \sys can be deployed using the \texttt{LD\_PRELOAD} trick~\cite{pulo2009fun}. This allows intercepting and overriding NCCL or RCCL symbols with the corresponding \sys implementation, enabling seamless integration with existing DL applications. Figure~\ref{fig:launch_system} illustrates a transparent, multi-vendor GPU cluster execution scenario. Only the necessary runtime and libraries for the GPU devices present on each node are required. \sys dynamically loads backend components (such as those for CUDA or HIP) as needed. \sys can be executed across multiple nodes using existing cluster execution utilities like \texttt{srun} from SLURM~\cite{yoo2003slurm} and \texttt{mpirun} for MPI~\cite{gabriel2004open}. 

\subsection{GPU-Aware Workload Balancing}

In a cluster of homogeneous GPUs, using the same micro-batch size for each GPU is a common practice. However, our setup involves heterogeneous GPUs, specifically NVIDIA and AMD GPUs that have varying performance capabilities. Simply assigning the same batch size to all GPUs can lead to significant underutilization of the faster devices, as training only continues after all GPUs have processed their respective micro-batches. As a result, the overall throughput is limited by the slowest GPU in the group. To address this issue, we adopt a straightforward yet effective strategy: we adjust the micro-batch size according to the GPU's compute capability~\cite {jia2022whale, lin2025kaitian}. 

Let $s_i$ denote the processing speed (tokens/s) of GPU $i$, obtained from brief offline profiling, and let $B$ be the total micro-batch size per step. We assign each GPU a micro-batch size $b_i$ proportional to its processing speed, i.e., $b_i = B \cdot \frac{s_i}{\sum_j s_j}$, with integer rounding. This ensures that all GPUs finish their local computation in approximately the same time, since $b_i / s_i$ is equalized across devices.

\section{Evaluation}
\label{sec:eval}

We evaluate \sys on a cluster comprising nodes equipped with GPUs from different vendors. We examine the communication performance and assess the overall performance of training DL models.

\subsection{Experimental Setup}

\paragraph{System configuration.}

We conduct experiments on a four-node cluster, consisting of two nodes equipped with four NVIDIA GPUs and two nodes with four AMD GPUs. This configuration allows us to evaluate communication performance across different GPU types (i.e., NVIDIA-to-AMD and vice versa). The detailed specifications of our system configuration are provided in Appendix~\ref{subsec:system-configurations}. Our results characterize cross-vendor training on this 16-GPU cluster, a scale comparable to the previous heterogeneous GPU systems summarized in Table~\ref{tab:hetero_training_scale}.

\paragraph{Comparison baselines.}

We compare \sys with state-of-the-art vendor-provided libraries, NCCL~\cite{nvidia_nccl} and RCCL~\cite{amd_rccl}, which also serve as core engines in existing CCLs~\cite{zhou2025extensible, si2025collective, shah2025msccl++, hidayetoglu2025hiccl}. To our knowledge, no prior work supports cross-vendor GPU communication; we adopt these as baselines due to their vendor-specific optimizations and seamless integration with modern DL frameworks. We further evaluate \sys against MPI in Appendix~\ref{subsec:mpi-comparison}.

\paragraph{Models and Implementation.}
To evaluate end-to-end training performance with \sys, we use LLMs from two widely used model families, GPT~\cite{radford2019language} and LLaMA~\cite{grattafiori2024llama}. For each family, we evaluate multiple model sizes, covering a wide range of memory and computation intensity scenarios.
Experiments are conducted using PyTorch~\cite{paszke2019pytorch} with DeepSpeed~\cite{rasley2020deepspeed} for parallelization. We provide the model details in Appendix~\ref{subsec:model-details} and implementation details in Appendix~\ref{subsec:implementation-details}. 

\subsection{Communication Performance}
\label{eval-comm-perf}

\paragraph{Point-to-point communication.}

\begin{wrapfigure}{r}{0.47\linewidth}
    \centering
    \vspace{-1em}
    \includegraphics[width=\linewidth]{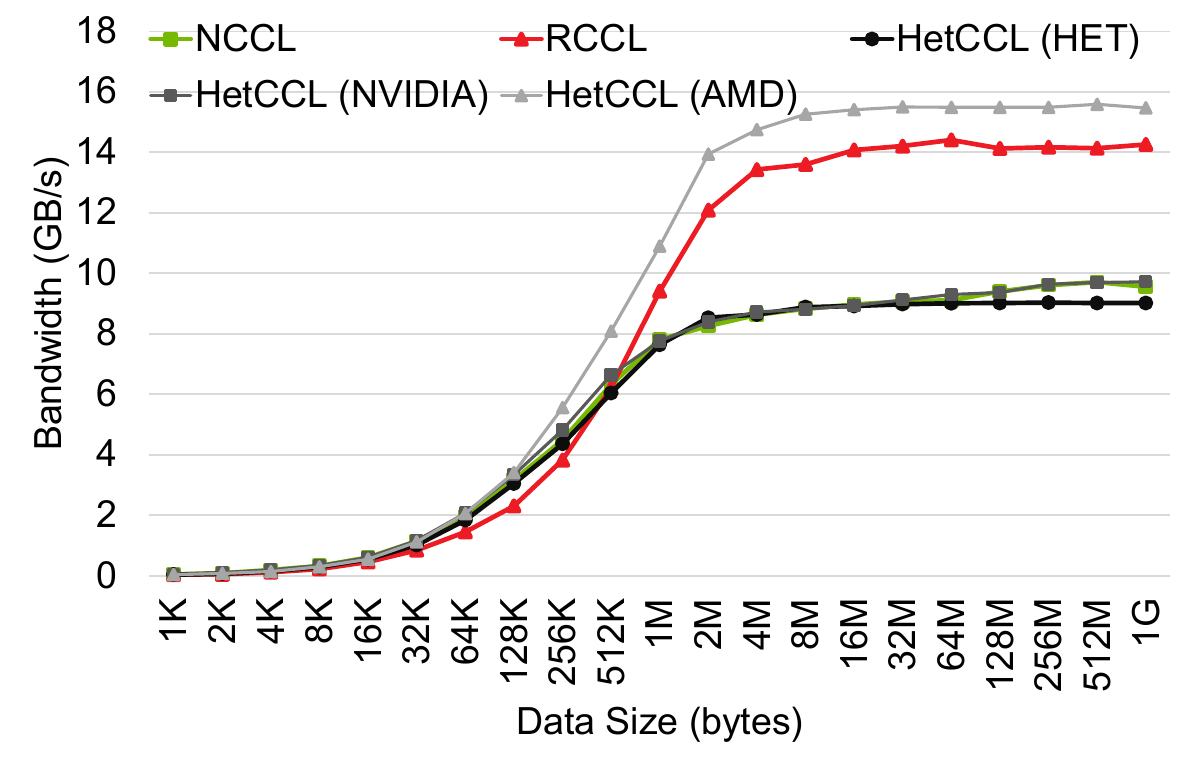}
    \caption{Performance of point-to-point communication with RDMA mechanism.}
    \label{fig:p2p_perf}
    \vspace{-1em}
\end{wrapfigure}

Figure~\ref{fig:p2p_perf} shows the RDMA-based point-to-point communication bandwidth across data transfer sizes from 1KB to 1GB. We evaluate three variants of \sys: \sys (NVIDIA) for NVIDIA-only, \sys (AMD) for AMD-only, and \sys (HET) for heterogeneous NVIDIA-AMD communication. Vendor-optimized NCCL and RCCL are used as upper-bound baselines for homogeneous settings, where the measured peak bandwidths are bounded by the underlying GPU interconnects (PCIe Gen3 on NVIDIA nodes and PCIe Gen4 on AMD nodes).

\sys (NVIDIA) achieves performance comparable to NCCL, indicating negligible overhead in homogeneous NVIDIA environments. \sys (AMD) slightly outperforms RCCL due to more effective configuration choices; we observed that RCCL’s default parameters are suboptimal on our system, and manual tuning (e.g., increasing \texttt{NCCL\_MAX\_NTHREADS}) narrows this gap. By integrating both NCCL and RCCL, \sys can selectively adopt configurations that improve overall performance.
In heterogeneous cross-vendor communication, \sys (HET) achieves performance close to \sys (NVIDIA) or NCCL, as overall bandwidth is bounded by the slower endpoint. This demonstrates that \sys effectively enables cross-vendor RDMA communication while approaching near-native performance in homogeneous configurations. We discuss the effect of RDMA in Appendix~\ref{subsec:ablation-rdma}.

\begin{figure*}[t]
    \centering
    \begin{subfigure}{0.329\textwidth}
        \centering
        \includegraphics[width=\linewidth]{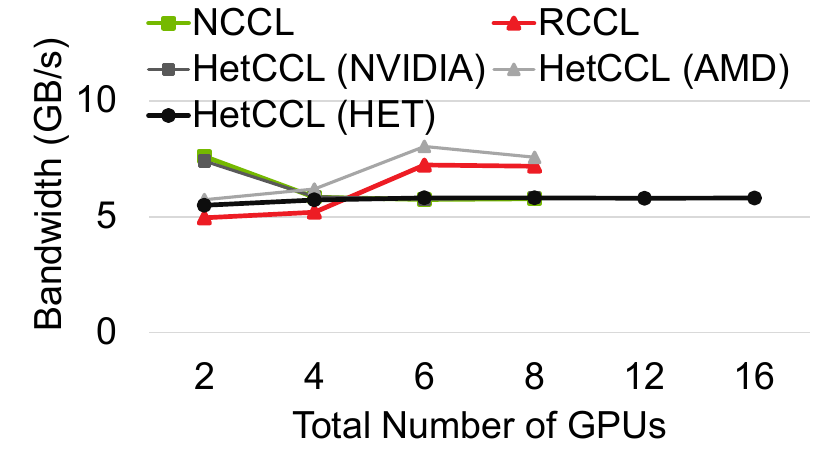}
        \caption{All-Reduce}
        \label{fig:ccl_allreduce}
    \end{subfigure}
    \begin{subfigure}{0.329\textwidth}
        \centering
        \includegraphics[width=\linewidth]{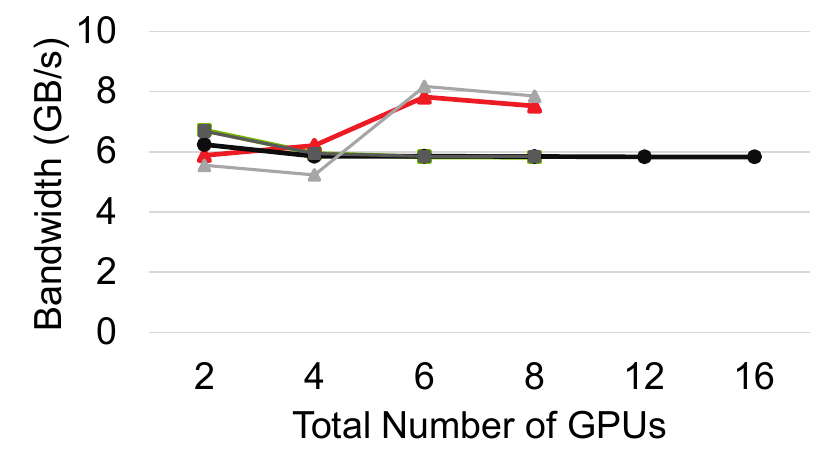}
        \caption{All-Gather}
        \label{fig:ccl_allgather}
    \end{subfigure}
    \begin{subfigure}{0.329\textwidth}
        \centering
        \includegraphics[width=\linewidth]{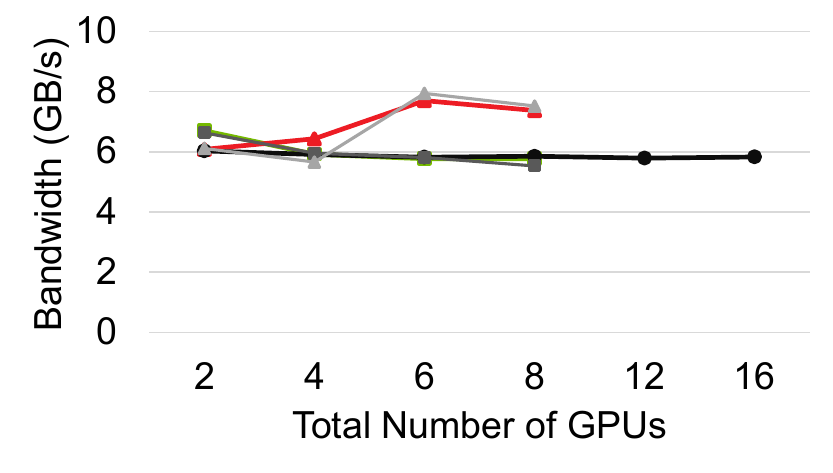}
        \caption{Reduce-Scatter}
        \label{fig:ccl_reducescatter}
    \end{subfigure}

    \vspace{0\baselineskip}
    
    \caption{Performance of representative collective communication operations.}
    \label{fig:ccl_perf}
\end{figure*}

\paragraph{Collective communication.}

We evaluate the collective communication performance of NCCL, RCCL, and \sys under three configurations: \sys (NVIDIA), \sys (AMD), and \sys (HET) for heterogeneous NVIDIA+AMD setups. We cover all widely used collective operations—All-Reduce, All-Gather, Reduce-Scatter, Reduce, Broadcast, and All-to-All—using a 1GB data size, as large message sizes are representative of training scenarios (e.g., gradient accumulation during backpropagation)~\cite{shah2025msccl++}. Figure~\ref{fig:ccl_perf} shows the achieved bandwidth of three collective operations; results for the remaining operations are reported in Appendix~\ref{subsec:additional-collectives}. The x-axis denotes the total number of GPUs participating in the collective operation. For homogeneous setups, we evaluate up to 8 GPUs across two nodes, using 1-4 GPUs per node, which reflects the maximum homogeneous configuration available in our cluster. In contrast, \sys (HET) enables cross-vendor communication and scales beyond this limit, reaching up to 16 GPUs across four nodes.

Across all operations, \sys (NVIDIA) achieves performance nearly identical to NCCL, indicating negligible overhead in homogeneous NVIDIA environments. \sys (AMD) performs comparably to or slightly better than RCCL due to more effective configuration choices. In heterogeneous settings, \sys (HET) consistently achieves performance bounded by the slower of the two vendor libraries. Depending on the scale and collective operation, either NCCL or RCCL becomes the bottleneck, which in turn limits \sys (HET). This behavior reflects \sys's design, which delegates communication to vendor-optimized backends for each GPU type. Moreover, \sys (HET) maintains stable bandwidth as the system scales to 12 and 16 GPUs, with 3 and 4 GPUs per node, respectively, demonstrating its efficiency and robustness in mixed-vendor environments.

\subsection{Training DL Models}
\label{eval-dl-training}

For efficient training, we use DeepSpeed ZeRO~\cite{rajbhandari2020zero} and evaluate both ZeRO-1 and ZeRO-3, which differ in how model states are partitioned and in the resulting collective communication patterns. We describe the partitioning and communication characteristics of each ZeRO stage in Appendix~\ref{subsec:comm-patterns-zero-stages}. Using initial profiling to quantify the NVIDIA-AMD GPU performance gap, \sys balances the load by proportionally assigning a larger micro-batch size to the faster GPUs. We discuss the effect of \sys's workload balancing on training performance as an ablation in Appendix~\ref{subsec:ablation-load-balancing}.

\begin{figure*}[t]
    \centering
    \captionsetup[subfigure]{skip=0pt} 
    \captionsetup{skip=5pt}            

    \begin{subfigure}[t]{0.98\textwidth}
        \centering
        \includegraphics[width=\linewidth]{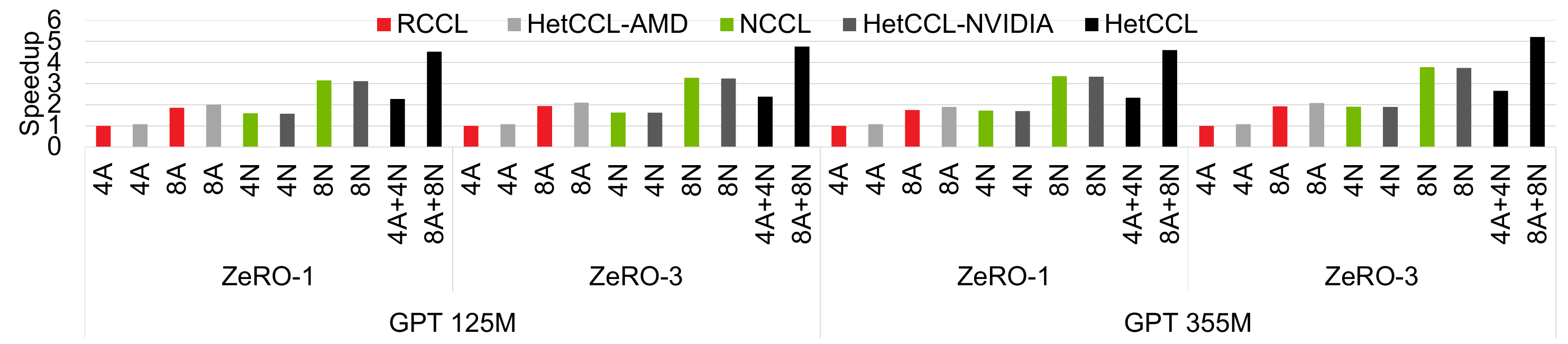}
        \caption{GPT}
        \label{fig:dl_perf_gpt}
    \end{subfigure}


    \begin{subfigure}[t]{0.98\textwidth}
        \centering
        \includegraphics[width=\linewidth]{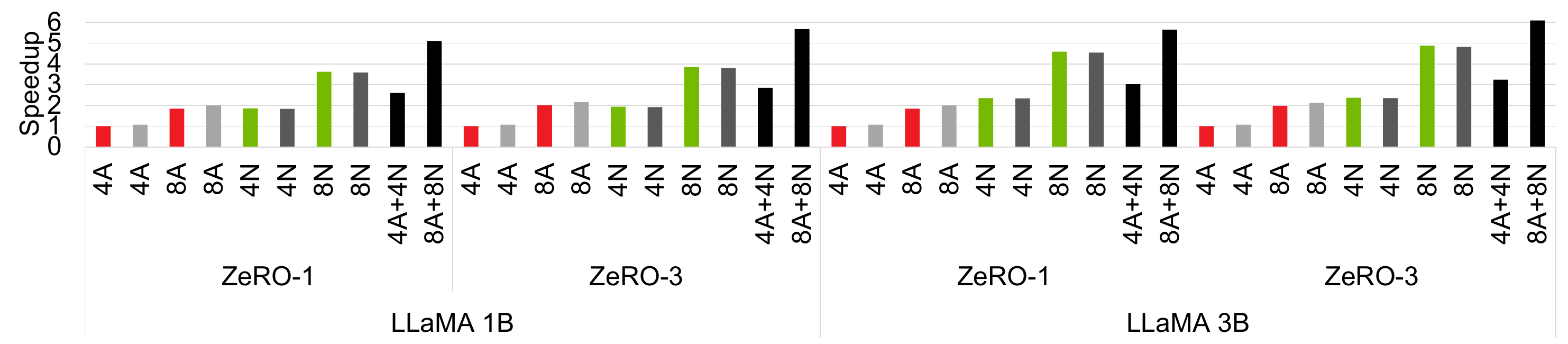}
        \caption{LLaMA}
        \label{fig:dl_perf_llama}
    \end{subfigure}

    \vspace{-0em} 

    \caption{LLM training speedup. 4N/8N or 4A/8A denotes homogeneous NVIDIA or AMD GPU training, while 4A+4N/8A+8N denotes heterogeneous NVIDIA+AMD GPU training.}
    \label{fig:dl_perf}
\end{figure*}

\paragraph{End-to-end Performance.}

Figure~\ref{fig:dl_perf} reports throughput speedup relative to RCCL on AMD GPUs. We evaluate homogeneous configurations using only AMD GPUs (4A, 8A) or only NVIDIA GPUs (4N, 8N), as well as heterogeneous configurations combining both vendors (4A+4N and 8A+8N). These setups span from single-node configurations with 4 GPUs to multi-node configurations with up to 16 GPUs, enabling us to evaluate scalability across both GPU and node counts.

\sys remains close to these native baselines on homogeneous GPUs. By jointly utilizing NVIDIA and AMD GPUs, \sys achieves average speedups of $1.38\times$ over NVIDIA-only (NCCL) training and $2.69\times$ over AMD-only (RCCL) training, demonstrating effective aggregation of heterogeneous compute resources with minimal overhead and reduced straggler effects. Here, we use a normalized metric, \textit{efficiency}, to quantify how closely heterogeneous GPU execution approaches the performance of homogeneous baselines. We define efficiency as the throughput of \sys's mixed-vendor execution (NVIDIA+AMD) normalized by the sum of throughputs from homogeneous NVIDIA and AMD runs. This ensures a fair comparison by comparing heterogeneous execution with homogeneous baselines using the same total number of GPUs.
Across all configurations, \sys achieves up to 97\% efficiency relative to the sum of homogeneous NCCL and RCCL throughputs, with workload balancing raising the efficiency for LLaMA-1B with ZeRO-3 from 89\% to this level and improving throughput by 1.09$\times$ (Table~\ref{tab:ablation-load-balancing}). Comparing ZeRO-1 and ZeRO-3 under identical configurations, we observe negligible differences in efficiency. \sys also achieves around 90\% efficiency, with no significant difference between GPT and LLaMA.

\sys demonstrates near-linear scalability when scaling from 8 GPUs (4A+4N) to 16 GPUs (8A+8N). Across all models, speedup nearly doubles with the number of GPUs, indicating that \sys scales effectively in multi-node, mixed-vendor GPU environments.  Unlike baselines limited to homogeneous GPUs, \sys enables multi-vendor GPUs to participate simultaneously, reducing time-to-solution and GPU-hour consumption, a common proxy for training cost. Additional results demonstrating \sys's compatibility with hybrid and expert parallelism are shown in Appendix~\ref{subsec:larger-scale-moe-models}.

\paragraph{Model Accuracy.}
To examine model output consistency in heterogeneous vendor GPU training, we analyze training convergence using LLaMA-1B, trained on the WikiText-103~\cite{merity2016pointer} dataset for 1K steps. We use BF16 precision and identical training configurations across NCCL (NVIDIA-only), RCCL (AMD-only), and \sys (NVIDIA+AMD) in this experiment.

We observe no discrepancies in loss across communication backends. The relative error of the final loss value across all pairwise comparisons among NCCL, RCCL, and \sys is below $7\times10^{-3}$, well within the numerical tolerance of BF16 (machine epsilon: $7.8125\times10^{-3}$). The training loss curves provided in Appendix~\ref{subsec:training-convergence} demonstrate closely overlapping convergence behavior and identical final model states. These results indicate that heterogeneous GPU training does not harm model integrity, and that \sys preserves model accuracy at a level consistent with NCCL and RCCL.
\section{Limitations and Future Work}
\label{sec:limitations}

While this work demonstrates that efficient training on a multi-vendor GPU cluster is possible with \sys, this section discusses the limitations of our study and directions for future work.

\paragraph{System Scope.}
\sys currently targets a system setting in which each node contains GPUs
from a single vendor, while the overall cluster consists of nodes with different vendors. We chose this scope because intra-node mixed-vendor GPU configurations are uncommon in practice~\cite{hu2024characterization, tang2025h2, zhu2025megascale}, and our current testbed also follows this deployment model. Accordingly, we did not include intra-node mixed-vendor experiments in this work. That said, the core mechanisms of \sys are not inherently limited to inter-node heterogeneity. The TACC abstraction and per-vendor device code compilation mechanism remain valid for heterogeneous GPUs within a node. Extending to intra-node multi-vendor environments would additionally require handling cross-vendor P2P/SHM paths, which we plan to explore as future work.


\paragraph{Dynamic Load Balancing.} 
\sys currently determines micro-batch ratios through an initial offline profiling phase, which is sufficient to achieve stable and effective load balancing, as confirmed in the evaluation results. As a complementary extension, we plan to explore online re-profiling that dynamically updates micro-batch ratios based on per-batch compute times, to further improve robustness against potential performance drift caused by factors such as thermal throttling and resource contention.

\section{Conclusion}

Training infrastructure is increasingly heterogeneous, yet today’s DL software ecosystem assumes a single GPU vendor, forcing costly modifications or leaving accelerators underutilized. \sys removes this barrier by enabling cross-vendor GPU communication while preserving existing vendor software stacks. Through RDMA-based communication and two key mechanisms—(1) runtime API abstraction and (2) kernel-level integration—\sys achieves near-native performance in homogeneous settings and operates efficiently in mixed-vendor environments. By enabling transparent training on heterogeneous vendor GPUs, \sys expands the feasible design space for ML practitioners, allowing larger batch sizes, higher training throughput, and more flexible use of available accelerators without modifying application code or compromising model correctness.

\begin{ack}
This work was partially supported by the National Research Foundation of Korea (NRF) under Grant No. RS-2023-00222663 (Center for Optimizing Hyperscale AI Models and Platforms) and under Grant No. A400-20260031, and by the Institute for Information and Communications Technology Promotion (IITP) under Grant No. 2018-0-00581 (CUDA Programming Environment for FPGA Clusters) and No. RS-2025-02304554 (Efficient and Scalable Framework for AI Heterogeneous Cluster Systems), all funded by the Ministry of Science and ICT (MSIT) of Korea. It was also partially supported by the Korea Health Industry Development Institute (KHIDI) under Grant No. RS-2025-25454559 (Frailty Risk Assessment and Intervention Leveraging Multimodal Intelligence for Networked Deployment in Community Care), funded by the Ministry of Health and Welfare (MOHW) of Korea. Additional support was provided by the BK21 Plus Program for Innovative Data Science Talent Education (Department of Data Science, Seoul National University, No. 5199990914569) and the BK21 FOUR Program for Intelligent Computing (Department of Computer Science and Engineering, Seoul National University, No. 4199990214639), both funded by the Ministry of Education (MOE) of Korea. This work was also partially supported by the Advanced GPU Utilization Support Program, funded by the Ministry of Science and ICT (MSIT) of Korea and operated by the National IT Industry Promotion Agency (NIPA). This work was also supported in part by Samsung Research, Samsung Electronics Co., Ltd. Research facilities were provided by the Institute of Computer Technology (ICT) at Seoul National University.
\end{ack}

\bibliographystyle{plainnat}
\bibliography{references}

\appendix
\newpage
\section*{Appendix}

The appendix is organized as follows:
\begin{enumerate}
    \item Appendix~\ref{sec:extended-related-work} discusses extended related work.
    \item Appendix~\ref{sec:tacc-api-functions} includes the runtime API functions implemented in TACC.
    \item Appendix~\ref{sec:experimental-details} provides experimental details. 
    \item Appendix~\ref{sec:additional-experimental-results} includes additional experimental results.
    \item Appendix~\ref{sec:ablation-study} provides an ablation study.
    \item Appendix~\ref{sec:impact-statement} includes the impact statement.

\end{enumerate}

We publicly release the code at: \url{https://github.com/mcrl/hetccl}.


\newpage

\section{Extended Related Work}
\label{sec:extended-related-work}

Our work relates broadly to various research areas in distributed training with heterogeneous GPUs. In this section, we discuss the communication patterns arising from parallelization in DL (Appendix~\ref{subsec:comm_patterns_in_dl}), and the initiatives aimed at enhancing the performance of DL model training or inference in heterogeneous GPU environments (Appendix~\ref{subsec:dl_on_hetero_gpus}).

\subsection{Parallelism and Communication Patterns in DL}
\label{subsec:comm_patterns_in_dl}

\paragraph{Parallelism in DL.} 
The rapid growth of DL, particularly LLMs, has driven an unprecedented increase in model scale—from hundreds of billions to trillions of parameters. To meet the corresponding rise in computational and memory demands, state-of-the-art DL frameworks~\cite{shoeybi2019megatron, narayanan2021efficient, korthikanti2023reducing, rajbhandari2020zero, ren2021zero-offload, rajbhandari2021zero-infinity} now rely heavily on multi-GPU execution through various forms of parallelism, including data, tensor, pipeline, and expert parallelism~\cite{rajbhandari2020zero, shoeybi2019megatron, huang2019gpipe, shazeer2017outrageously}. These parallelization strategies inevitably introduce intensive communication between GPUs. Each parallelization strategy requires specific types of collective communication operations. 

\paragraph{Communication patterns in DL.}
\textit{\textbf{Fully sharded data parallelism (FSDP)}}~\cite{rajbhandari2020zero} partitions model parameters across GPUs and performs training by gathering all parameters during the forward pass using All-Gather, and distributing gradients during the backward pass via Reduce-Scatter collective communication. \textit{\textbf{Tensor parallelism}}~\cite{shoeybi2019megatron} splits individual weight matrices across GPUs. It computes partial results in parallel, which are then aggregated using All-Reduce collective communication to obtain the final result. \textit{\textbf{Pipeline parallelism}}~\cite{huang2019gpipe} involves sequential point-to-point transfers between pipeline stages. \textit{\textbf{Expert parallelism}}~\cite{shazeer2017outrageously, li2023accelerating, fedus2022switch} for Mixture-of-Experts (MoE) architectures demands complex All-to-All collective communication across GPU devices. During inference, \textit{\textbf{Prefill-decode disaggregation}}~\cite{zhong2024distserve} requires low-latency point-to-point communication to minimize user-perceived latency.

As models grow larger and more complex, the demand for GPUs in both training and inference continues to rise. To meet this scaling pressure, many organizations expand their clusters incrementally, often ending up with heterogeneous GPU environments that mix vendors such as NVIDIA and AMD. This trend, however, poses a critical challenge: enabling efficient communication across GPUs from different vendors. Traditional communication libraries and hardware interconnects are optimized for homogeneous setups and cannot support or exploit cross-vendor communication.

\subsection{DL on Heterogeneous GPUs}
\label{subsec:dl_on_hetero_gpus}
Modern GPU clusters typically consist of devices with different compute capabilities, memory sizes, and interconnect bandwidths due to upgrades over time or hardware reuse. Effectively utilizing these heterogeneous resources for DL model training has become a significant focus of research. Many existing approaches explore various combinations of parallelism and resource allocation strategies to optimize the training of DL models on these diverse GPUs. These efforts can be broadly categorized into two groups: (1) automatic parallelization, which plans how a single job runs in parallel across different devices, and (2) cluster-level schedulers, which determine how jobs are assigned to various hardware resources over time.

\begin{table}[ht]
\centering
\caption{Heterogeneity type and evaluation scale of representative related systems.}
\vspace{0.7ex}
\label{tab:hetero_training_scale}
\resizebox{0.94\linewidth}{!}{
    \begin{tabular}{lccl}
    \toprule
    \textbf{System} & \textbf{Vendor Setting} & \textbf{Max \#GPUs} & \textbf{GPU Configuration} \\
    \midrule
    \addlinespace[1pt]
    HeteroG~\cite{yi2020optimizing} & Single-vendor & 12 & NVIDIA V100, P100, GTX 1080Ti \\
    HetSeq~\cite{ding2021hetseq} & Single-vendor & 16 & NVIDIA Titan Xp, GTX 1080Ti, P100 \\
    HetPipe~\cite{park2020hetpipe} & Single-vendor & 16 & NVIDIA TITAN V, TITAN RTX, RTX 2060, Quadro P4000 \\
    AMP~\cite{li2022amp} & Single-vendor & 16 & NVIDIA V100, T4 \\
    Metis~\cite{um2024metis} & Single-vendor & 32 & NVIDIA P100, T4, V100 \\
    \midrule
    \sys & Cross-vendor & 16 & NVIDIA V100, AMD Radeon Pro W7800 \\
    \bottomrule
    \end{tabular}
}
\end{table}

\paragraph{Automatic parallelization on heterogeneous GPUs.}
Several systems are designed to automatically determine parallelism strategies that take into account hardware heterogeneity within a job. 
HeteroG~\cite{yi2020optimizing} proposes a Graph Neural Network (GNN)-based optimization framework that selects hybrid parallelism strategies (i.e., data and model parallelism) along with communication methods for heterogeneous GPUs in a cluster. 
HetSeq~\cite{ding2021hetseq} enables parallel training across GPUs with varying performance levels by allowing different batch sizes for each device and synchronizing gradients through weighted averaging, although it focuses exclusively on data parallelism.
HetPipe~\cite{park2020hetpipe} combines pipeline parallelism among virtual workers with data parallelism across them, introducing a synchronization model to ensure convergence even when using lower-tier GPUs.
AMP~\cite{li2022amp} formulates a search space for degrees of parallelism and pipeline assignments, utilizing a heterogeneity-aware cost model to identify efficient strategies through dynamic programming. 
Metis~\cite{um2024metis} expands on this approach by incorporating device grouping and asymmetric workload balancing. It employs a heterogeneity-aware search algorithm that prunes inefficient configurations and adjusts intra-stage parallelism based on device capacity. 

Table~\ref{tab:hetero_training_scale} summarizes the types of heterogeneity and evaluation scales for these systems. Consistent with these prior heterogeneous GPU systems, \sys targets representative mid-scale clusters formed through incremental GPU integration, while further addressing cross-vendor interoperability between NVIDIA and AMD GPUs. However, these previous studies focus on parallelization strategies for heterogeneous GPUs, whereas we concentrate on implementing a collective communication library for heterogeneous GPUs.

\paragraph{Heterogeneity-aware cluster scheduling.}
In addition to efforts focused on parallelization, another area of research involves job-level scheduling for heterogeneous GPUs in a cluster. Gavel~\cite{narayanan2020heterogeneity} schedules jobs based on throughput profiling across different GPU types, aiming to optimize job completion times while adhering to fairness constraints. However, it does not allow for the splitting of a job across devices of different types. Sia~\cite{jayaram2023sia} supports elastic and adaptive DL jobs by dynamically scaling their resource usage over time. It also considers heterogeneity during resource allocation, but still limits each job to a single type of device.

\begin{figure*}[ht]
  \centering
  \begin{minipage}{0.92\textwidth}
    \centering
    \rule{\linewidth}{0.5pt}

    \begin{minipage}{0.48\linewidth}
\begin{lstlisting}[style=tacc]
taccGetAvailablePlatforms();
taccSetPlatform();
taccSetPlatformAuto();
taccGetPlatform();
taccThreadExchangeStreamCM();
taccMalloc();
taccMallocHost();
taccMallocManaged();
taccHostAlloc();
taccGetDevice();
taccSetDevice();
taccGetErrorString();
taccGetLastError();
taccExtMallocWithFlags();
taccDeviceCanAccessPeer();
taccFreeHost();
taccDeviceGetAttribute();
taccStreamCreateWithFlags();
taccMemcpy();
taccMemcpyAsync();
taccFree();
taccMemset();
taccMemsetAsync();
taccStreamQuery();
taccStreamSynchronize();
taccStreamDestroy();
taccEventCreate();
taccEventCreateWithFlags();
taccEventQuery();
taccEventRecord();
taccStreamWaitEvent();
taccEventDestroy();
taccLaunchHostFunc();
\end{lstlisting}
    \end{minipage}\hfill
    \begin{minipage}{0.48\linewidth}
\begin{lstlisting}[style=tacc,firstnumber=34]
taccIpcOpenMemHandle();
taccIpcCloseMemHandle();
taccIpcGetMemHandle();
taccHostRegister();
taccHostUnregister();
taccHostGetDevicePointer();
taccDeviceGetPCIBusId();
taccGetDeviceProperties();
taccDeviceGetByPCIBusId();
taccDeviceEnablePeerAccess();
taccGetDeviceCount();
taccFuncSetAttribute();
taccFuncGetAttributes();
taccExtLaunchKernel();
taccLaunchKernel();
taccDeviceSetLimit();
taccStreamGetCaptureInfo_v2();
taccGraphAddHostNode();
taccGraphInstantiate();
taccGraphLaunch();
taccGraphExecDestroy();
taccGraphDestroy();
taccDeviceSynchronize();
taccStreamBeginCapture();
taccStreamEndCapture();
taccDeviceReset();
taccPointerGetAttributes();
taccExtStreamCreateWithCUMask();
taccStreamCreate();
taccExtGetLinkTypeAndHopCount();
taccHostMalloc();
taccHostFree();
taccEventElapsedTime();
\end{lstlisting}
    \end{minipage}

    \rule{\linewidth}{0.5pt}
  \end{minipage}
  \caption{TACC API functions.}
  \label{fig:tacc-runtime-api}
\end{figure*}

While these systems acknowledge GPU heterogeneity in terms of architectural generations or performance variability, they generally operate under the assumption that all GPUs come from a single vendor. This is due to the lack of support for collective communication across different vendors. Consequently, the heterogeneity they address is somewhat limited, which helps keep system complexity manageable. Enabling true heterogeneity—where GPUs from different vendors can participate in a single collective communication operation—could open up new research avenues in both parallelization and scheduling.


\section{TACC API functions}
\label{sec:tacc-api-functions}

Figure~\ref{fig:tacc-runtime-api} shows the full list of functions implemented in \sys's runtime API abstraction layer, TACC. TACC abstracts the CUDA and HIP runtime APIs used in NCCL and RCCL.


\section{Experimental Details}
\label{sec:experimental-details}

This section describes the experimental details, including the system configuration of our four-node cluster (Appendix~\ref{subsec:system-configurations}), details of DL models used in our experiments (Appendix~\ref{subsec:model-details}), implementation details of DL frameworks for LLM training (Appendix~\ref{subsec:implementation-details}), and the communication patterns that arise when training with DeepSpeed ZeRO (Appendix~\ref{subsec:comm-patterns-zero-stages}).

\subsection{System Configuration}
\label{subsec:system-configurations}

\begin{table*}[ht]
\centering
\caption{Hardware configuration of the four-node multi-vendor GPU cluster.}
\vspace{-1.0ex}
\label{tab:cluster_spec}
\resizebox{0.80\linewidth}{!}{
\begin{tabular}{|l|c|c|}
\hline
 & \textbf{NVIDIA (2 nodes)} & \textbf{AMD (2 nodes)} \\
\hline \hline
Mainboard & GIGABYTE MZ52-G40 & ASRock ROMED8-2T \\
\hline
CPU & 1 $\times$ AMD EPYC 7452 32-Core & 2 $\times$ AMD EPYC 7313 16-Core \\
\hline
Memory & 8 $\times$ DDR4-2666 64GB & 4 $\times$ DDR4-3200 16GB \\
\hline
GPU & 4 $\times$ NVIDIA Tesla V100-PCIe & 4 $\times$ AMD Radeon Pro W7800 \\
\hline
NIC & Mellanox ConnectX-6 InfiniBand HDR & Mellanox ConnectX-6 InfiniBand HDR \\
\hline
GPU PCIe & Gen3 & Gen4 \\
\hline
GPU Driver & 550.54.15 (CUDA 12.4) & 6.12.12 (ROCm 6.4.0) \\
\hline
\end{tabular}
}
\end{table*}

To comprehensively evaluate \sys, we configure a heterogeneous GPU cluster comprising multiple NVIDIA and AMD GPU nodes connected with a uniform InfiniBand interconnect. All nodes use the same InfiniBand NIC and fabric. This setup enables us to compare the results of cross-vendor GPU communication with those from homogeneous GPU environments, where GPUs are housed within the same vendor's nodes. The detailed system configurations are described in Table~\ref{tab:cluster_spec}. 

Even though the experiments were carried out on a PCIe cluster without high-speed GPU interconnects like NVLink~\cite{nvidia_nvlink} or xGMI~\cite{amd_xgmi}, the proposed method and findings are still relevant to more advanced environments. This is because vendor libraries like NCCL and RCCL are designed to automatically leverage available hardware features when present. \sys is built on top of these libraries and naturally inherits such functionality. To demonstrate that \sys remains effective even in the presence of high-bandwidth interconnects, we include an evaluation on a single-node high-end GPU system with high-bandwidth interconnects in Appendix~\ref{subsec:perf-high-end-system}.

\subsection{Models}
\label{subsec:model-details}

\begin{table}[ht]
\centering
\caption{Specifications of models used in the experiments.}
\vspace{0.8ex}
\resizebox{0.50\linewidth}{!}{
\begin{tabular}{@{}lccc@{}}
\toprule
\textbf{Model} & \textbf{Parameter size} & \textbf{Sequence length} & \textbf{Vocab size} \\
\midrule
\addlinespace[1pt]
\midrule
\multirow{2}{*}{GPT}   & 125M & \multirow{2}{*}{1024}  & \multirow{2}{*}{50257} \\
                       & 355M &                        &                         \\
\midrule
\multirow{2}{*}{LLaMA} & 1B   & \multirow{2}{*}{8192}  & \multirow{2}{*}{32000} \\
                       & 3B   &                        &                         \\
\bottomrule
\end{tabular}
\label{tab:model_config}
}
\end{table}

Table~\ref{tab:model_config} shows the details of the DL models used in Section~\ref{eval-dl-training}. We follow the remaining hyperparameter configurations (e.g., number of layers, hidden size, attention heads) from the original papers~\cite{achiam2023gpt, grattafiori2024llama}. For each model, we choose the maximum feasible batch size before GPU out-of-memory (OOM) errors. Experiments on larger-scale or MoE model training are in Appendix~\ref{subsec:larger-scale-moe-models}.

\subsection{Implementation Details}
\label{subsec:implementation-details}

\paragraph{Tunable Parameters in CCLs.} NCCL and RCCL hard-code a number of tunable parameters whose defaults are not optimal on every system, so some adjustment is expected when porting to new hardware. Two parameters are not aligned between the two: HetCCL (AMD) is built with \texttt{NCHANNELS\_PER\_NET\_PEER} raised from 1 to 2 and \texttt{MAX\_NTHREADS} from 256 to 512, because RCCL's defaults left significant bandwidth unused on our cluster, whereas we ran RCCL as released in order to reproduce it faithfully. However, we did not intend this as a performance claim as written in Section~\ref{eval-comm-perf}, and recompiling RCCL with the same two constants closes the gap entirely. All other tunables are aligned across the three implementations: \texttt{NCCL\_BUFFSIZE}, \texttt{NCCL\_MIN/MAX\_NCHANNELS}, \texttt{NCCL\_STEPS}, and all \texttt{NCCL\_IB\_*} settings, with \texttt{NCCL\_ALGO=RING} applied uniformly to every run.

\paragraph{Training DL Models.} We use PyTorch~\cite{paszke2019pytorch} as our DL framework and employ the AdamW~\cite{loshchilov2017decoupled} optimizer, incorporating activation checkpointing~\cite{chen2016training} to reduce memory usage. To maximize computational efficiency and reduce memory footprint, we use FP16 precision for all end-to-end performance evaluations in Section~\ref{eval-dl-training}. For model convergence and numerical consistency analysis, however, we follow the training precision used in the original work~\cite{grattafiori2024llama} to ensure faithful reproduction of model convergence behavior and therefore conduct training using BF16. For parallelization, we adopt DeepSpeed~\cite{rasley2020deepspeed}, a widely recognized DL optimization library developed by Microsoft. DeepSpeed is designed to be hardware-agnostic and currently supports accelerators from multiple vendors, including Intel, Huawei, NVIDIA, and AMD, which makes it a natural choice for heterogeneous multi-GPU training. We intentionally disable all forms of CPU or NVMe offloading~\cite{ren2021zero-offload, rajbhandari2021zero-infinity} in DeepSpeed to avoid confounding effects from host memory capacity differences in our cluster.

\subsection{Communication Patterns under Different ZeRO Stages}
\label{subsec:comm-patterns-zero-stages}

The ZeRO~\cite{rajbhandari2020zero} family of optimizations in DeepSpeed~\cite{rasley2020deepspeed} includes several stages, each based on different components being partitioned. As a result, different ZeRO stages involve various collective communication operations, as shown in Table~\ref{tab:zero_comm}. We use both ZeRO-1 and ZeRO-3 in our experiments to evaluate various collective communication patterns. ZeRO-1 primarily employs All-Gather for optimizer states and All-Reduce for gradients,
while ZeRO-3 utilizes All-Gather for parameters and Reduce-Scatter for gradients.

\begin{table}[ht] 
\centering 
\caption{Communication patterns across DeepSpeed ZeRO stages.} \vspace{0.7ex} 
\label{tab:zero_comm} 
\resizebox{1.0\linewidth}{!}{ 
    \begin{tabular}{lccc} 
    \toprule \textbf{ZeRO Stage} & \textbf{Stage 1} & \textbf{Stage 2} & \textbf{Stage 3 (FSDP)} \\ 
    \midrule 
    \addlinespace[1pt] 
    \midrule 
    Optimizer State (OS) & Sharded & Sharded & Sharded \\
    Gradient (G) & Replicated & Sharded & Sharded \\ 
    Parameter (P) & Replicated & Replicated & Sharded \\ 
    \midrule 
    Communication Ops & All-Gather (OS), All-Reduce (G) & All-Gather (P), Reduce-Scatter (G) & All-Gather (P), Reduce-Scatter (G) \\ 
\bottomrule 
\end{tabular} 
} 
\end{table}


\section{Additional Experimental Results}
\label{sec:additional-experimental-results}

This section provides additional experimental results, including the performance of other collective communication operations (Appendix~\ref{subsec:additional-collectives}), loss convergence of LLM training (Appendix~\ref{subsec:training-convergence}), a comparison with different vendor library versions (Appendix~\ref{subsec:recent-nccl-rccl-comparison}), model training with hybrid and expert parallelism (Appendix~\ref{subsec:larger-scale-moe-models}), a performance comparison against GPU-aware MPI (Appendix~\ref{subsec:mpi-comparison}), a performance evaluation on high-end GPU systems (Appendix~\ref{subsec:perf-high-end-system}), and a run-to-run statistical variability (Appendix~\ref{subsec:statistical_variability}).

\subsection{Other Collective Communication Operations}
\label{subsec:additional-collectives}

\begin{figure*}[ht]
    \centering
    \begin{subfigure}{0.329\textwidth}
        \centering
        \includegraphics[width=\linewidth]{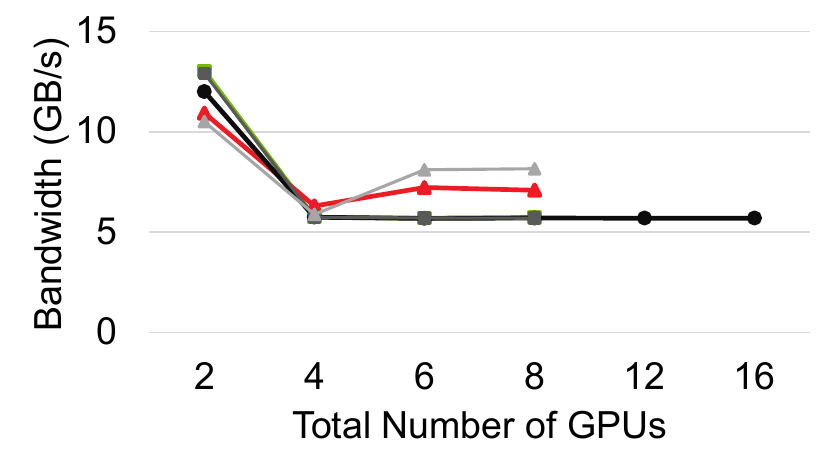}
        \caption{Reduce}
        \label{fig:ccl_reduce}
    \end{subfigure}
    \begin{subfigure}{0.329\textwidth}
        \centering
        \includegraphics[width=\linewidth]{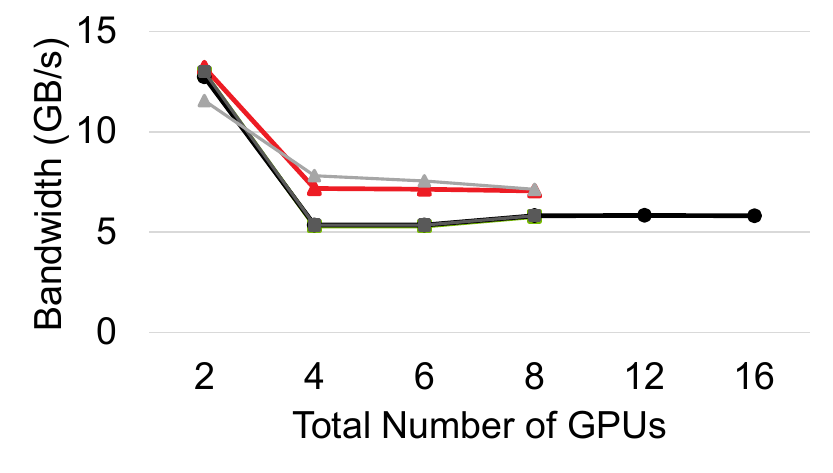}
        \caption{Broadcast}
        \label{fig:ccl_broadcast}
    \end{subfigure}
    \begin{subfigure}{0.329\textwidth}
        \centering
        \includegraphics[width=\linewidth]{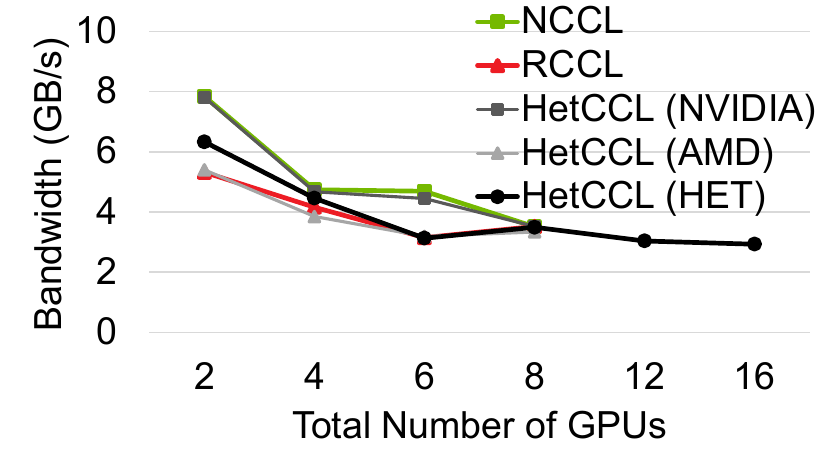}
        \caption{All-to-All}
        \label{fig:ccl_alltoall}
    \end{subfigure}
    \caption{Performance of other representative collective communication operations.
}
    \label{fig:additional_collectives}
\end{figure*}

We provide the performance of Reduce, Broadcast, and All-to-All collective communication operations in Figure~\ref{fig:additional_collectives}. The behavior is consistent across various collective patterns as discussed in Section~\ref{eval-comm-perf}, reflecting \sys's RDMA-based communication and internal strategy of delegating communication to vendor libraries: NCCL for NVIDIA GPUs and RCCL for AMD GPUs.

\subsection{LLM Training Convergence}
\label{subsec:training-convergence}

\begin{figure*}[ht]
    \centering
    \begin{subfigure}{0.49\textwidth}
        \centering
        \includegraphics[width=\linewidth]{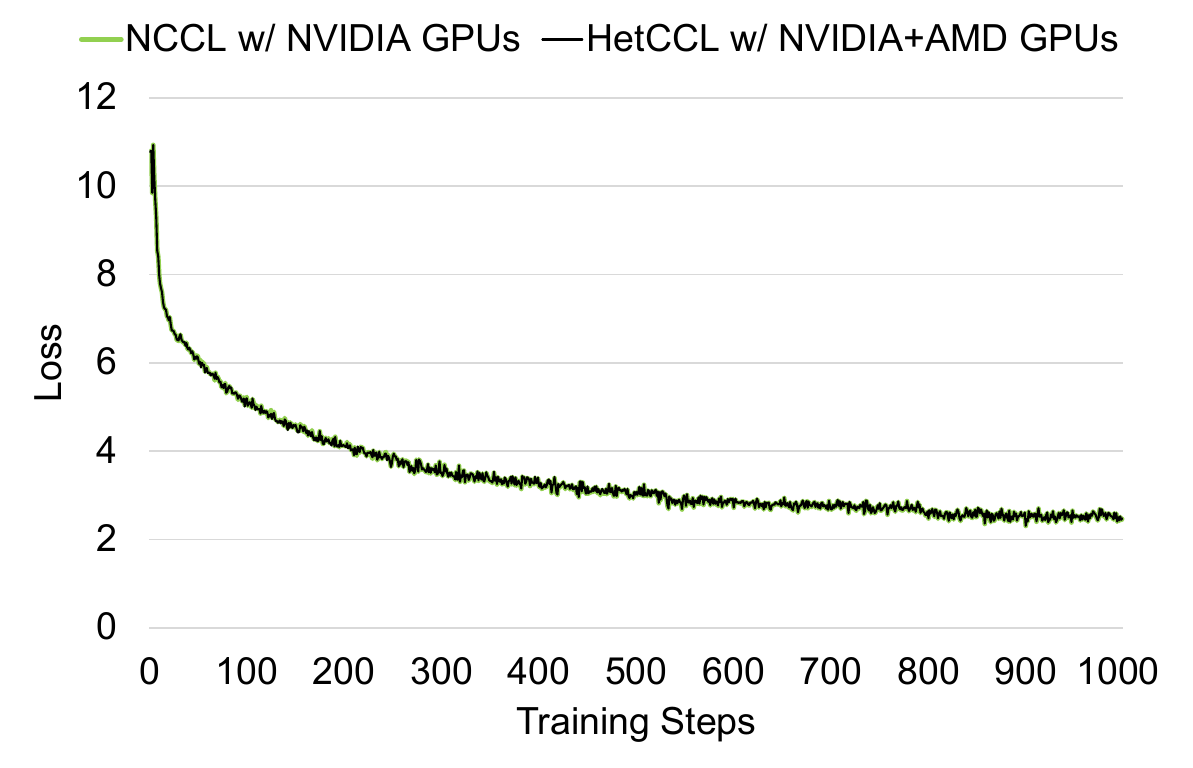}
        \caption{NCCL vs. \sys}
        \label{fig:training_convergence_nccl}
    \end{subfigure}
    \begin{subfigure}{0.49\textwidth}
        \centering
        \includegraphics[width=\linewidth]{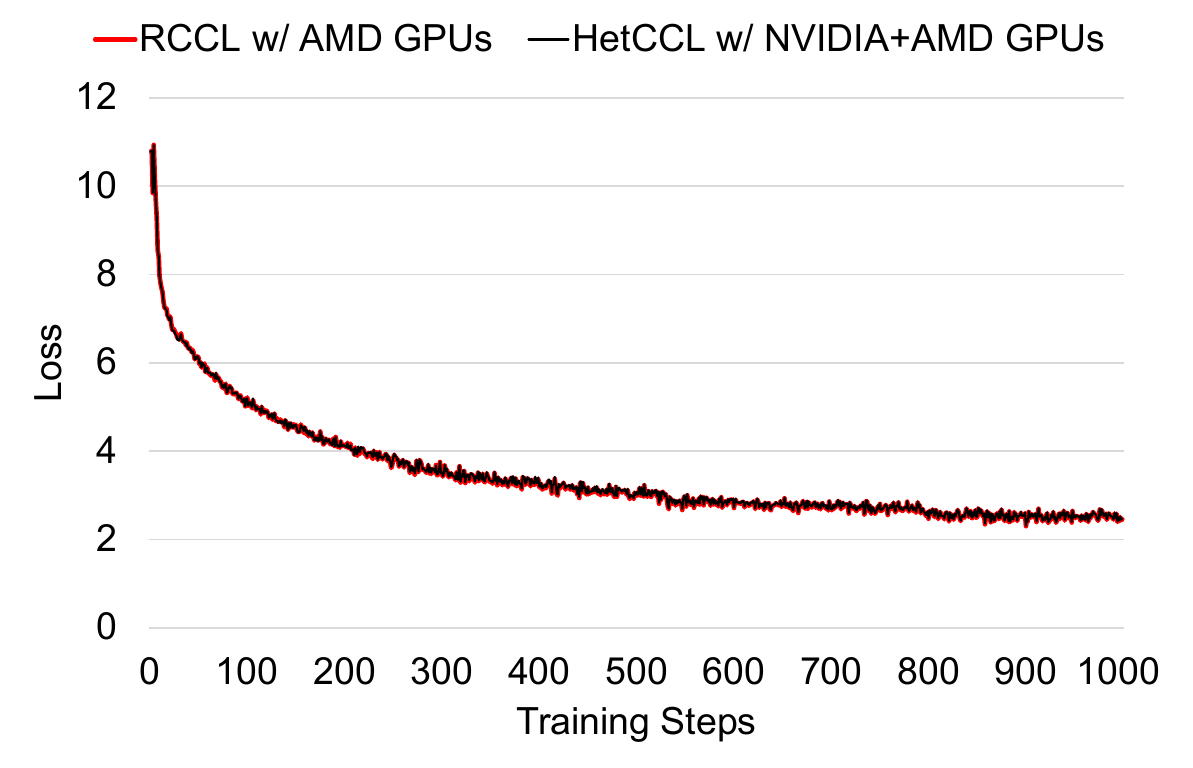}
        \caption{RCCL vs. \sys}
        \label{fig:training_convergence_rccl}
    \end{subfigure}
    \caption{Training loss convergence of LLaMA-1B on WikiText-103.}
    \label{fig:training_convergence}
\end{figure*}

We evaluate training convergence to verify that heterogeneous multi-vendor GPU training with \sys does not introduce model accuracy degradation or numerical instability. Figure~\ref{fig:training_convergence} compares the training loss curves of LLaMA-1B across homogeneous and heterogeneous configurations. We observe no noticeable divergence in convergence behavior or final loss values when using \sys compared to vendor-specific baselines (i.e., NCCL or RCCL). Table~\ref{tab:wikitext_ppl} further shows that \sys achieves validation and test perplexity comparable to NCCL and RCCL, with only small relative differences (within 1.0\%) from NCCL.

\begin{table}[ht]
\centering
\caption{Validation and test perplexity of LLaMA-1B on WikiText-103.}
\vspace{0.8ex}
\label{tab:wikitext_ppl}
\resizebox{0.58\linewidth}{!}{
    \begin{tabular}{lcc}
    \toprule
    \textbf{CCL (GPU Vendor)} & \textbf{Validation PPL} & \textbf{Test PPL} \\
    \midrule
    NCCL (NVIDIA GPUs) & 15.70 & 16.30 \\
    RCCL (AMD GPUs) & 15.54 & 16.15 \\
    \sys (NVIDIA + AMD GPUs) & 15.77 & 16.47 \\
    \bottomrule
    \end{tabular}
    }
\end{table}
 
This is because the implementation and execution behaviors of DL frameworks are consistent across NVIDIA and AMD systems. In addition, NVIDIA CUDA and AMD ROCm kernels exhibit numerically close behavior. Furthermore, NCCL and RCCL collectives are consistent within expected floating-point tolerances. Therefore, despite differences in floating-point implementations and reduction ordering across vendors, under deterministic settings (i.e., identical random seeds and initial weights without non-deterministic operations such as dropout), LLM training on hybrid NVIDIA--AMD systems exhibits stable behavior consistent with homogeneous training.

\subsection{Comparison between Different Versions of Vendor Libraries}
\label{subsec:recent-nccl-rccl-comparison}

\begin{figure}[ht]
    \centering
    \begin{subfigure}{0.329\textwidth}
        \centering
        \includegraphics[width=\linewidth]{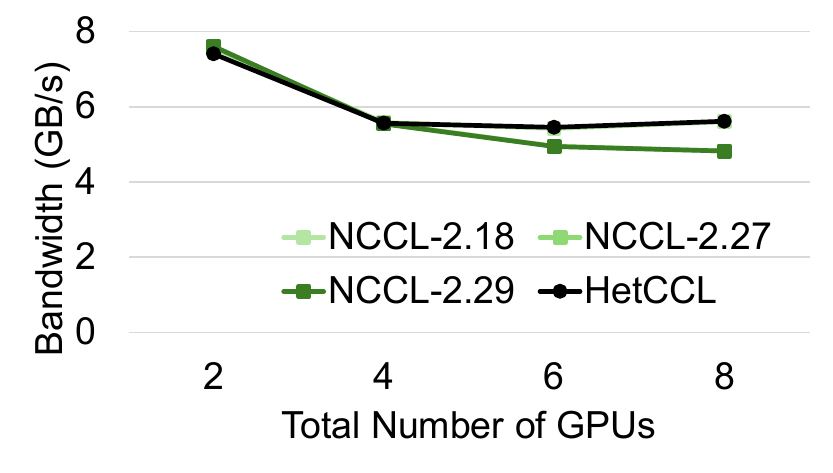}
        \caption{All-Reduce}
        \label{fig:ccl_allreduce}
    \end{subfigure}
    \begin{subfigure}{0.329\textwidth}
        \centering
        \includegraphics[width=\linewidth]{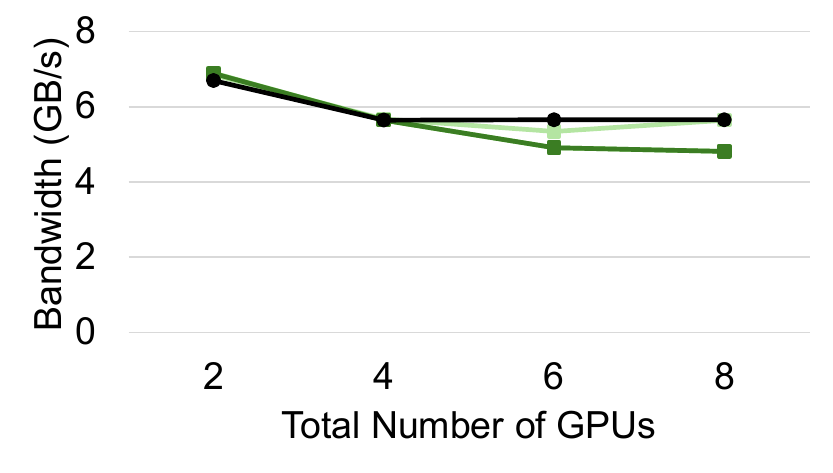}
        \caption{All-Gather}
        \label{fig:ccl_allgather}
    \end{subfigure}
    \begin{subfigure}{0.329\textwidth}
        \centering
        \includegraphics[width=\linewidth]{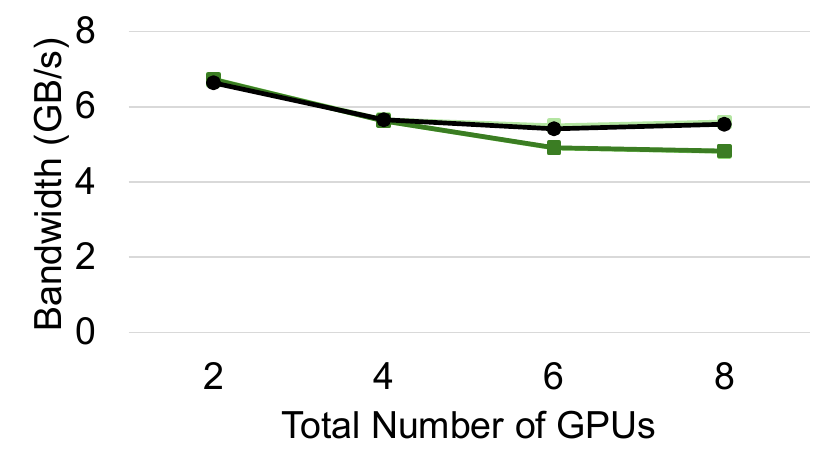}
        \caption{Reduce-Scatter}
        \label{fig:ccl_reducescatter}
    \end{subfigure}

    \caption{Collective communication performance across different NCCL versions.}
    \label{fig:nccl-version-ccl}
\end{figure}


We evaluate whether the choice of vendor library version affects the validity of our baseline comparison.
In the main evaluation, we use NCCL version 2.18.3 and the corresponding RCCL version 6.0.2 as vendor-optimized baselines, since \sys has been implemented on top of these versions.
To confirm that this does not underestimate recent versions of vendor libraries, we compare \sys against newer NCCL releases, including NCCL v2.27.7 and v2.29.7, for both point-to-point and collective communication.

\begin{wrapfigure}{r}{0.49\linewidth}
    \centering
    \vspace{-1em}
    \includegraphics[width=\linewidth]{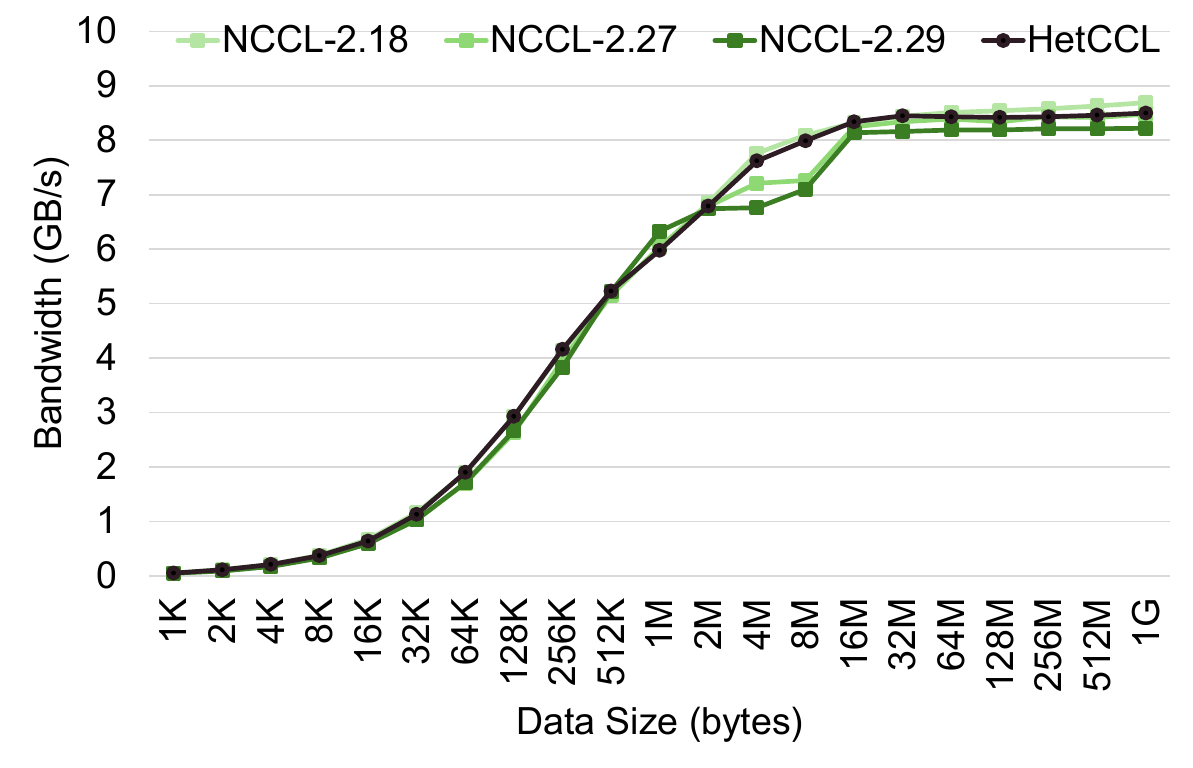}
    \caption{Point-to-point communication performance across different NCCL versions.}
    \label{fig:nccl-version-p2p}
    \vspace{-1em}
\end{wrapfigure}

As shown in Figure~\ref{fig:nccl-version-ccl} and Figure~\ref{fig:nccl-version-p2p}, the performance differences across NCCL versions are nearly negligible overall. This is consistent with the general development pattern of NCCL, where new releases primarily add support for newer hardware generations (e.g., NVSwitch, Blackwell architectures) and bug fixes, rather than substantially altering performance on existing hardware. We observe a similar trend for RCCL, which is expected because RCCL is maintained as a fork of NCCL and incorporates updates from every NCCL release into its codebase.
In some cases, especially for larger message sizes or larger GPU counts, newer NCCL versions are slightly slower than NCCL v2.18.3. We note that the goal of this work is not to analyze version-specific performance differences within NCCL or RCCL.
Rather, our focus is to demonstrate that \sys can provide cross-vendor communication while preserving near-native performance on homogeneous platforms.
We also note that since \sys delegates vendor-local communication to NCCL and RCCL through its runtime abstraction layer, newer NCCL and RCCL releases can be integrated with minimal changes.


\subsection{Training Models with Hybrid and Expert Parallelism}
\label{subsec:larger-scale-moe-models}

We also evaluate \sys on larger-scale models with hybrid parallelism and sparse MoE models with expert parallelism.
While the main experiments in Section~\ref{eval-dl-training} focus on data-parallel-only training to isolate \sys's core communication efficiency, this setting limits the maximum model size by per-GPU memory capacity.
To evaluate larger models, we use Megatron-DeepSpeed~\cite{megatron_deepspeed}, combining ZeRO-based data parallelism~\cite{rajbhandari2020zero} with Megatron tensor parallelism~\cite{shoeybi2019megatron}.

\begin{table}[ht]
    \centering
    \caption{LLaMA 7B and 13B model training with Megatron-DeepSpeed.}
    \label{tab:larger-model-results}
    \vspace{0.7ex}
    \resizebox{0.91\linewidth}{!}{
        \begin{tabular}{lcccc}
            \toprule
            \multirow{2}{*}{CCL (\# of GPUs)} 
                & \multicolumn{2}{c}{LLaMA 7B} 
                & \multicolumn{2}{c}{LLaMA 13B} \\
            \cmidrule(lr){2-3} \cmidrule(lr){4-5}
                & Throughput (tokens/s) & Speedup 
                & Throughput (tokens/s) & Speedup \\
            \midrule
            RCCL (8xAMD)                  & 691  & 1.00 & 386  & 1.00 \\
            NCCL (8xNVIDIA)              & 1289 & 1.87 & 716  & 1.85 \\
            \sys (4xAMD + 4xNVIDIA)    & 932  & 1.35 & 537  & 1.39 \\
            \sys (8xAMD + 8xNVIDIA)    & 1806 & 2.61 & 1021 & 2.64 \\
            \bottomrule
        \end{tabular}
    }
\end{table}

For LLaMA 7B and 13B training, we use hybrid parallelism with TP degree 4 and ZeRO-3, where the DP degree is 2 for 8-GPU runs and 4 for 16-GPU runs. Table~\ref{tab:larger-model-results} reports the training throughput of LLaMA 7B/13B models with the same configuration (e.g., sequence length, vocabulary size) as in Table~\ref{tab:model_config}. 
Even at these larger scales, \sys maintains high efficiency, achieving 91\% efficiency for LLaMA 7B and 93\% efficiency for LLaMA 13B. These results show that \sys is not limited to the model sizes evaluated in the main paper and can scale to substantially larger dense LLMs when combined with standard parallel training techniques.

\begin{table}[ht]
    \centering
    \caption{Sparse MoE model training with expert parallelism.}
    \label{tab:moe-model-results}
    \vspace{0.7ex}
    \resizebox{0.61\linewidth}{!}{
        \begin{tabular}{lcc}
            \toprule
            CCL (\# of GPUs) & Throughput (tokens/s) & Speedup \\
            \midrule
            RCCL (8xAMD)               & 2792 & 1.00 \\
            NCCL (8xNVIDIA)            & 4287 & 1.54 \\
            \sys (4xAMD + 4xNVIDIA)  & 3435 & 1.23 \\
            \sys (8xAMD + 8xNVIDIA)  & 6441 & 2.31 \\
            \bottomrule
        \end{tabular}
    }
\end{table}

To train sparse MoE models, we use the MoE example script~\cite{megatron_deepspeed_moe} provided by Megatron-DeepSpeed, which implements GPT-style MoE models with expert parallelism.
As shown in Table~\ref{tab:moe-model-results}, \sys achieves 91\% efficiency on GPT-3 XL MoE with 8 experts and 4.1B total parameters.
This result further demonstrates that \sys is compatible not only with dense LLM training but also with sparse expert-parallel workloads.

We intentionally exclude extremely large models such as LLaMA 70B from this evaluation because they require CPU or NVMe offloading mechanisms~\cite{ren2021zero-offload, rajbhandari2020zero}.
Although technically feasible, such offloading introduces additional system-level factors, making it difficult to isolate the contribution of \sys's cross-vendor communication.


\subsection{Performance Comparison against GPU-Aware MPI}
\label{subsec:mpi-comparison}

Modern MPI implementations support GPU-aware communication~\cite{wang2013gpu}, allowing GPU device buffers to be passed directly to MPI primitives without explicit host staging.
In practice, GPU-aware MPI support is provided through vendor-specific backends, such as CUDA-aware MPI~\cite{cuda_aware_mpi} for NVIDIA GPUs and ROCm-aware MPI~\cite{rocm_aware_mpi} for AMD GPUs, enabling efficient communication on each respective platform~\cite{panda2021mvapich, chen2023mpi}.

OpenMPI uses the same host-side MPI library on NVIDIA and AMD nodes and loads the appropriate accelerator plugin on each node. Our benchmark application allocates GPU buffers through CUDA or HIP, so we compile it separately for each vendor. We launch both executables within a single MPI job, mapping ranks to the corresponding nodes. MPI itself is unmodified, and all communication uses standard GPU-aware MPI interfaces.

\begin{wrapfigure}{r}{0.49\linewidth}
    \centering
    \vspace{-1em}
    \includegraphics[width=\linewidth]{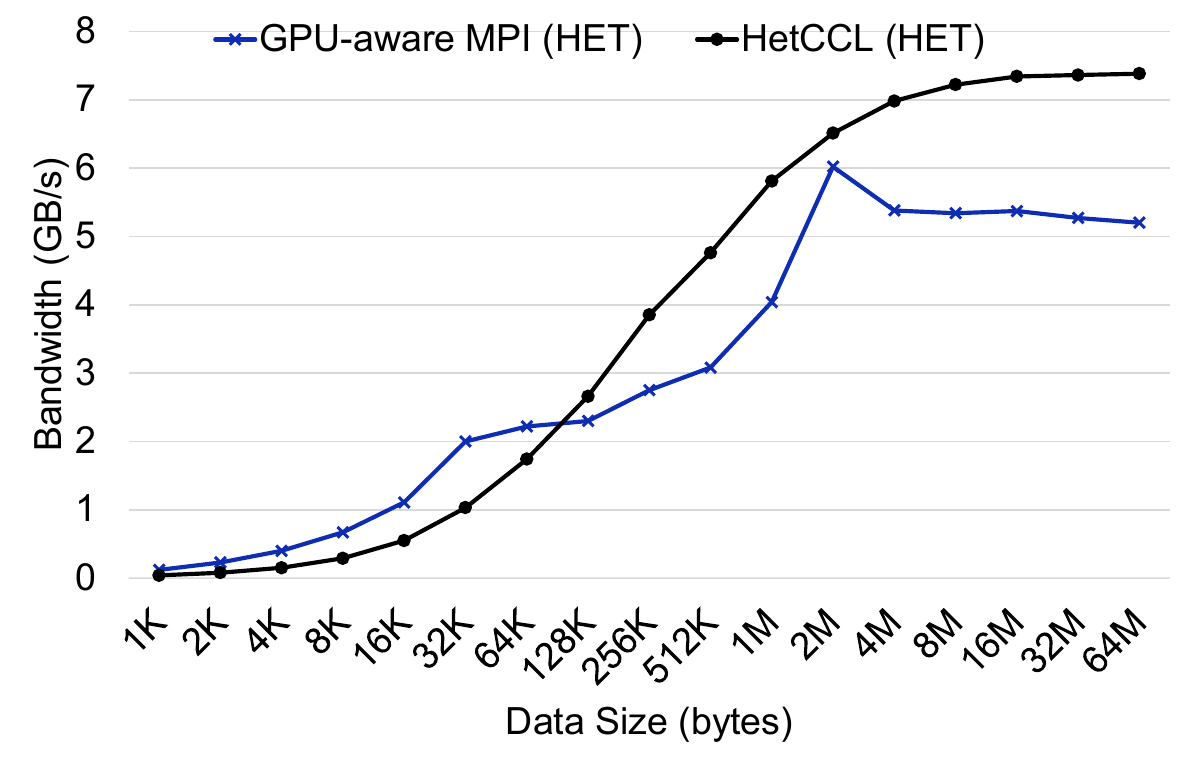}
    \caption{Point-to-point communication performance: GPU-aware MPI vs. \sys.}
    \label{fig:mpi_p2p_perf}
    \vspace{-1em}
\end{wrapfigure}

\begin{figure*}[ht]
    \centering
    \begin{subfigure}{0.49\textwidth}
        \centering
        \includegraphics[width=\linewidth]{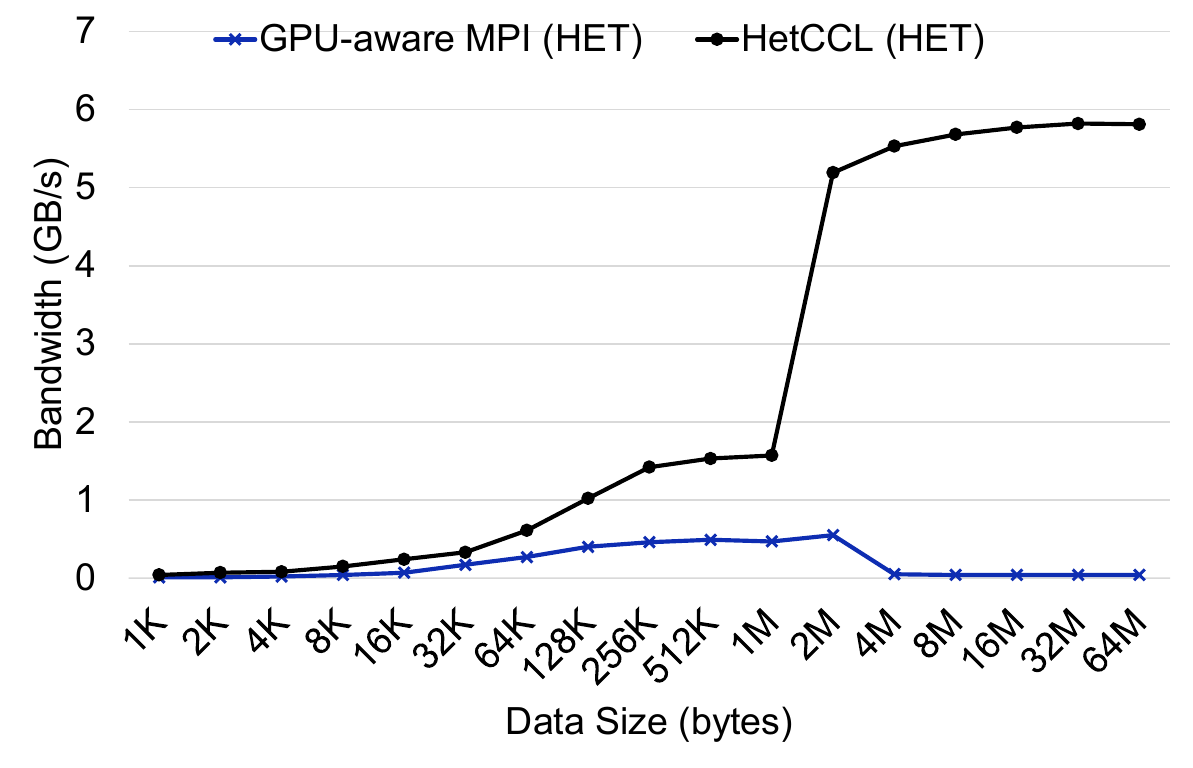}
        \caption{All-Reduce}
        \label{fig:mpi_allreduce}
    \end{subfigure}
    \begin{subfigure}{0.49\textwidth}
        \centering
        \includegraphics[width=\linewidth]{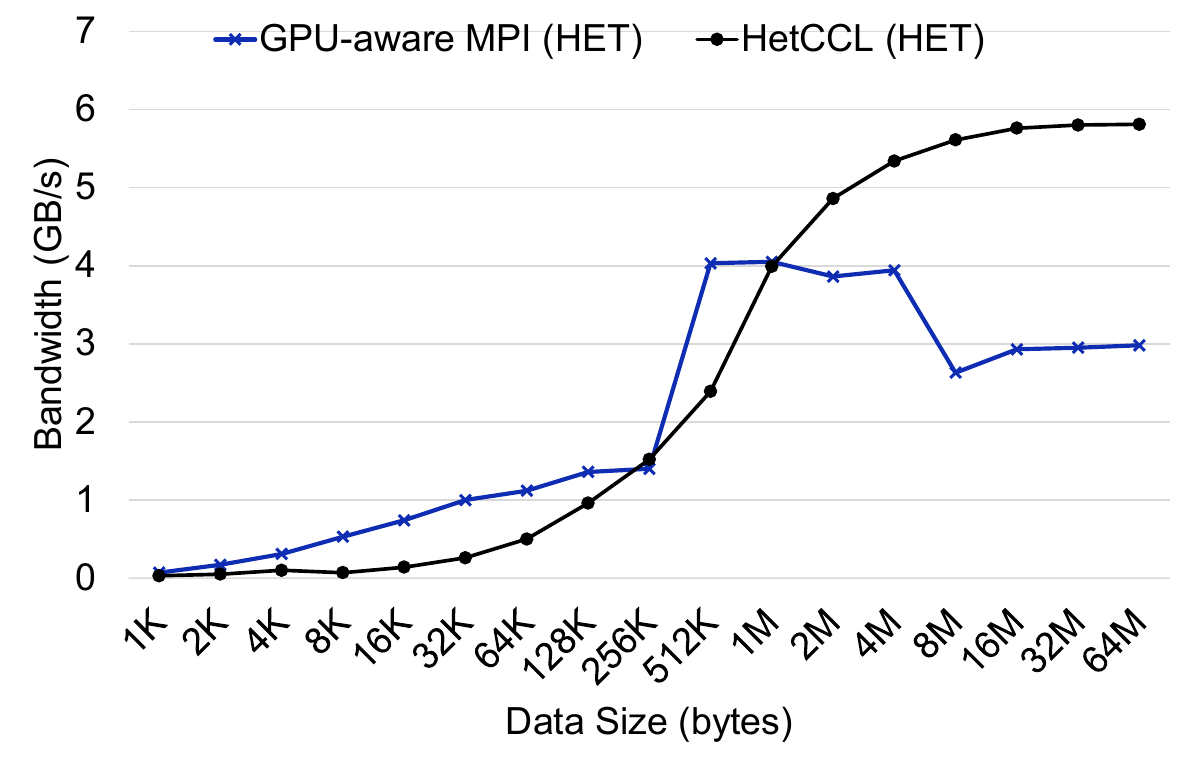}
        \caption{All-Gather}
        \label{fig:mpi_allgather}
    \end{subfigure}
    \caption{Collective communication performance on 16 GPUs: GPU-aware MPI vs. \sys.}
    \label{fig:mpi_ccl_perf}
\end{figure*}

In this evaluation, we implement heterogeneous GPU-aware MPI using OpenMPI (v5.0.2)~\cite{gabriel2004open} with UCX (v1.19.0)~\cite{shamis2015ucx}, one of the most widely used MPI stacks for accelerator-aware communication. 
We compare the performance of \sys against this baseline in both point-to-point and collective communication scenarios to assess the efficiency of the proposed design under heterogeneous GPU configurations.
Figure~\ref{fig:mpi_p2p_perf} presents the performance of cross-vendor point-to-point communication between NVIDIA and AMD GPUs, while Figure~\ref{fig:mpi_ccl_perf} reports the performance of All-Reduce and All-Gather collective operations on 16 GPUs spanning four nodes, where two nodes are equipped with NVIDIA GPUs and the remaining two nodes with AMD GPUs.

The results show that, consistent with prior studies~\cite{chen2023mpi, lin2025kaitian}, MPI achieves lower latency for small message sizes, whereas \sys delivers higher effective bandwidth for larger messages, making it more suitable for large-scale DL workloads dominated by large tensor transfers. This trend is especially evident in collective operations. Note that \sys outperforms GPU-aware MPI in All-Reduce across most of the evaluated message sizes, as MPI implementations typically do not natively provide GPU kernels and often incur additional overhead from host device staging for CPU-based reduction operations.
In contrast, \sys performs reductions entirely on the GPU, better exploiting the memory hierarchy and interconnect bandwidth of modern accelerators. Motivated by these limitations, recent studies~\cite{venkata2024unified} have explored the integration of optimized GPU kernels directly into MPI implementations to improve collective performance on accelerator-based systems. For end-to-end LLaMA-1B training on 16 GPUs with ZeRO-1 and a sequence length of 2048, \sys achieves an $11.4\times$ speedup over GPU-aware MPI, primarily due to lower communication overhead.

\subsection{Performance on High-End GPU System with High-Bandwidth Interconnect}
\label{subsec:perf-high-end-system}

\begin{figure}[ht]
    \centering
    
    \begin{subfigure}{0.49\textwidth}
        \centering
        \includegraphics[width=\linewidth]{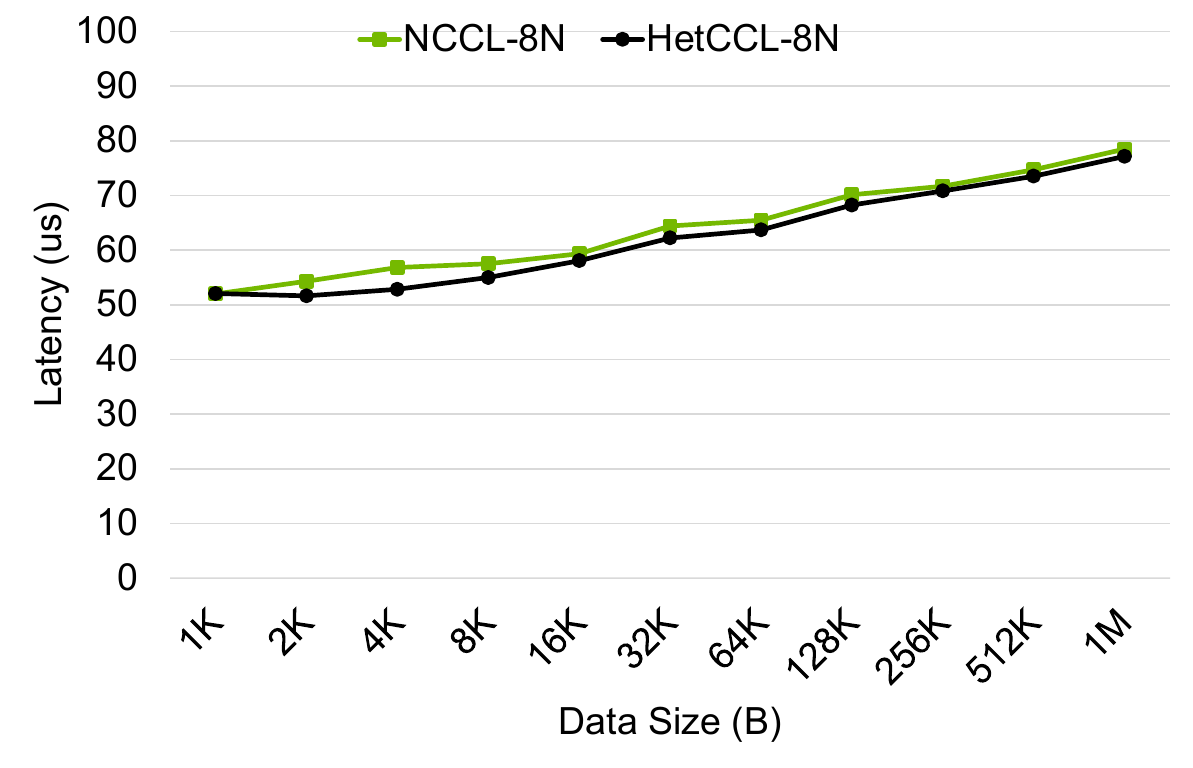}
        \caption{All-Reduce, 8$\times$H100 GPUs (1KB--1MB)}
        \label{fig:h100_lat}
    \end{subfigure}
    \hfill
    \begin{subfigure}{0.49\textwidth}
        \centering
        \includegraphics[width=\linewidth]{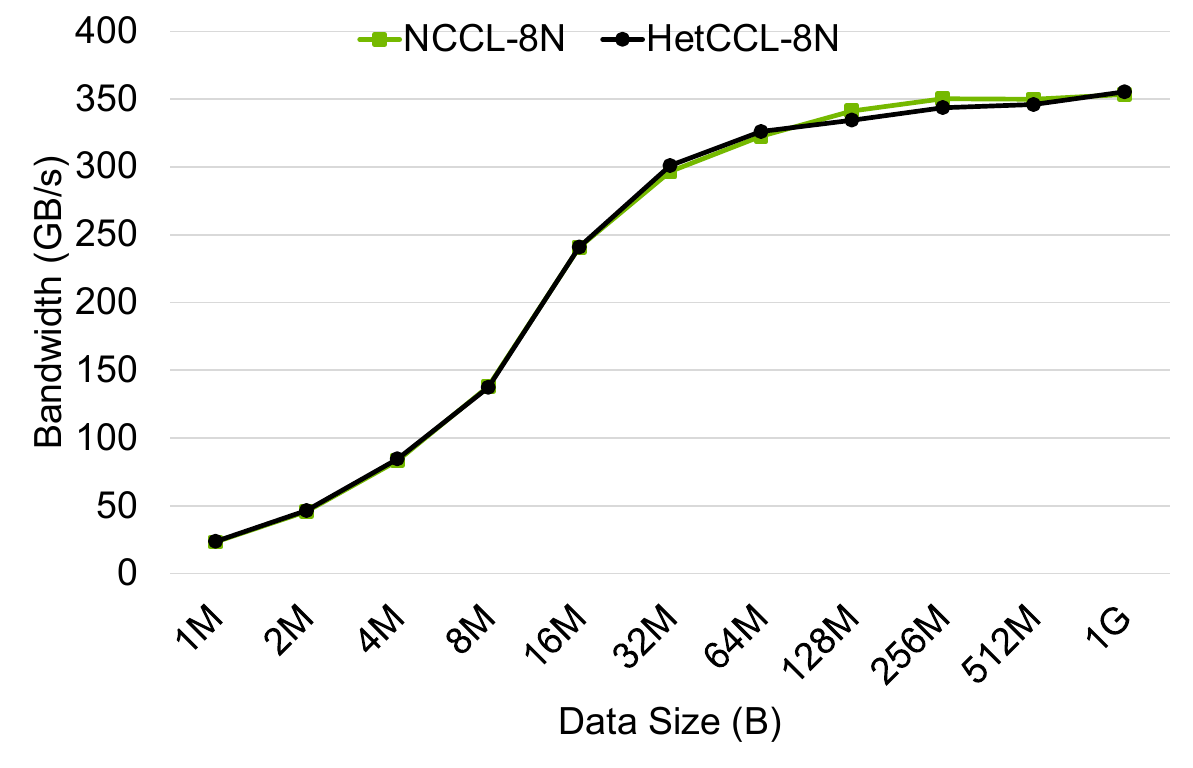}
        \caption{All-Reduce, 8$\times$H100 GPUs (1MB--1GB)}
        \label{fig:h100_bw}
    \end{subfigure}

    \vspace{0.3cm}
    
    \begin{subfigure}{0.49\textwidth}
        \centering
        \includegraphics[width=\linewidth]{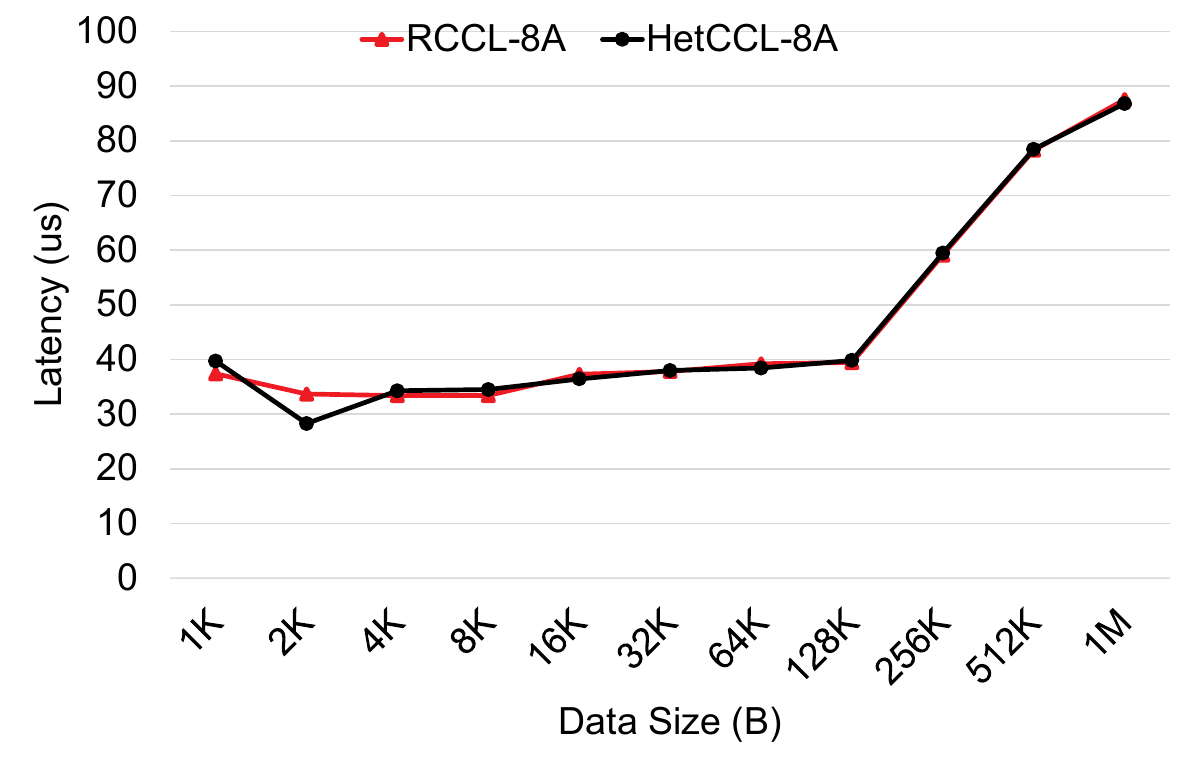}
        \caption{All-Reduce, 8$\times$MI300X GPUs (1KB--1MB)}
        \label{fig:mi300x_lat}
    \end{subfigure}
    \hfill
    \begin{subfigure}{0.49\textwidth}
        \centering
        \includegraphics[width=\linewidth]{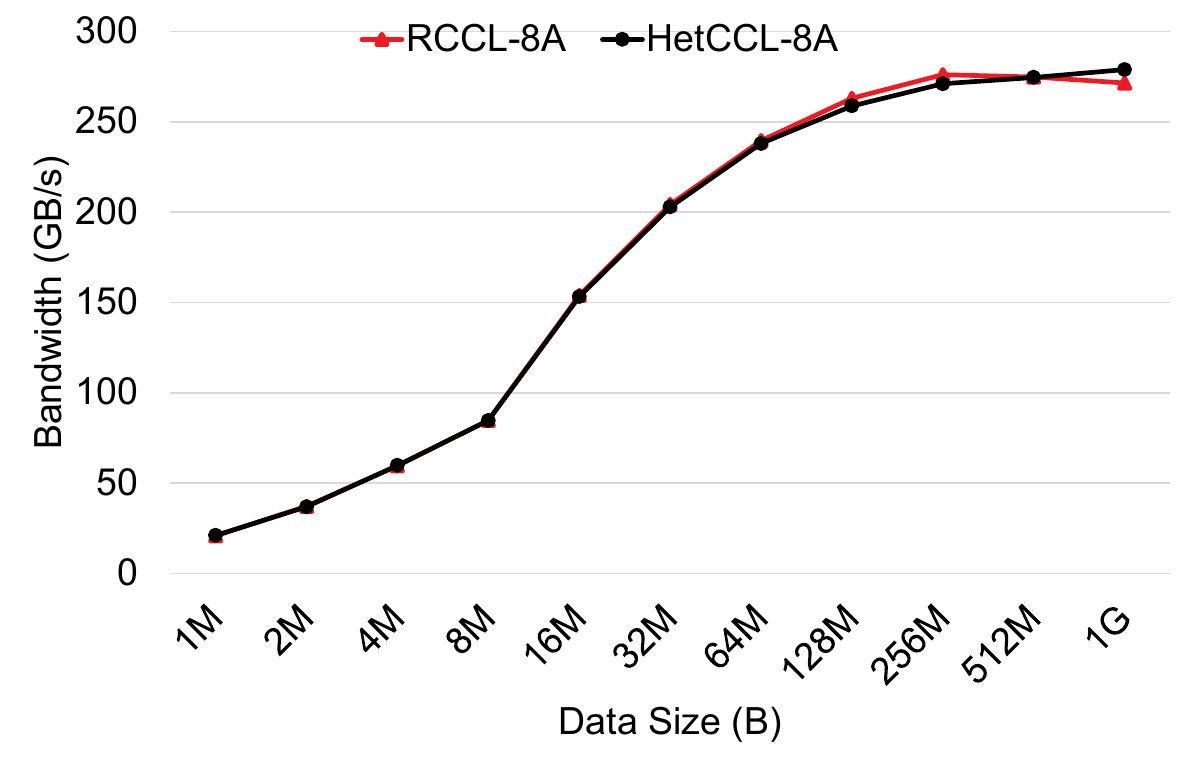}
        \caption{All-Reduce, 8$\times$MI300X GPUs (1MB--1GB)}
        \label{fig:mi300x_bw}
    \end{subfigure}

    \caption{Single node All-Reduce performance on NVIDIA and AMD high-end GPU systems.}
    \label{fig:high_end_eval}
\end{figure}

Due to the limited availability of a multi-vendor cluster environment equipped with the latest generation of accelerators, we indirectly validate the performance and extensibility of \sys through single-node experiments on homogeneous high-end GPU systems: an 8$\times$NVIDIA H100 system and an 8$\times$AMD MI300X system. Unlike our primary multi-vendor cluster (Table~\ref{tab:cluster_spec}), which relies on PCIe-based communication, these systems are equipped with high-bandwidth GPU interconnects, namely NVIDIA NVLink~\cite{nvidia_nvlink} and AMD xGMI~\cite{amd_xgmi}, providing significantly higher intra-node throughput. Note that even without a dedicated GPU-to-GPU interconnect (e.g., NVLink, xGMI), GPUs can still perform peer-to-peer (P2P) transfers over PCIe in our system setup.

We conduct single-node experiments to verify whether \sys introduces any performance overhead when leveraging vendor-specific high-speed interconnects. We compare \sys against NCCL on the H100 system and RCCL on the MI300X system using the All-Reduce collective operation with message sizes ranging from 1KB to 1GB. We separate the range of message sizes into small (1KB--1MB, presented in latency) and large (1MB--1GB, presented in bandwidth) for visibility.

As shown in Figure~\ref{fig:high_end_eval}, \sys matches the performance of vendor-optimized libraries on both platforms across all message sizes. While these results are obtained in homogeneous settings, they serve as a critical proxy for heterogeneous performance; by proving that \sys does not degrade the performance of the fastest available intra-node interconnects, we demonstrate that the library's abstraction layer is lightweight enough to support large-scale heterogeneous clusters without becoming a performance bottleneck.

\subsection{Statistical Variability}
\label{subsec:statistical_variability}



We repeat the collective communication configurations reported
in Table~\ref{tab:collective_variability} using three independent
job launches and report the mean $\pm$ standard deviation.
These measurements use a 1\,GB message size, at which bandwidth
has saturated.

\begin{table}[ht]
    \centering
    \caption{Run-to-run variability in collective communication bandwidth across two nodes.
    Values are mean $\pm$ standard deviation over three
    independent runs, in GB/s.}
    \label{tab:collective_variability}
    \vspace{0.7ex}
    \resizebox{0.92\linewidth}{!}{%
        \begin{tabular}{lccccc}
            \toprule
            Operation & NCCL & RCCL
            & HetCCL (NVIDIA) & HetCCL (AMD) & HetCCL (HET) \\
            \midrule
            All-Reduce
            & $5.80 \pm 0.02$ & $7.17 \pm 0.02$
            & $5.80 \pm 0.02$ & $7.85 \pm 0.03$
            & $5.81 \pm 0.02$ \\
            All-Gather
            & $5.83 \pm 0.01$ & $7.37 \pm 0.02$
            & $5.82 \pm 0.03$ & $7.55 \pm 0.04$
            & $5.84 \pm 0.02$ \\
            Reduce-Scatter
            & $5.85 \pm 0.02$ & $7.35 \pm 0.03$
            & $5.84 \pm 0.02$ & $7.91 \pm 0.03$
            & $5.84 \pm 0.02$ \\
            \bottomrule
        \end{tabular}%
    }
\end{table}

\begin{table}[ht]
    \centering
    \caption{Run-to-run variability in end-to-end training throughput for LLaMA-1B
    with ZeRO-3 and FP16. Values are mean $\pm$ standard
    deviation over three independent runs.}
    \label{tab:training_variability}
    \vspace{0.7ex}
    \resizebox{0.53\linewidth}{!}{%
        \begin{tabular}{lc}
            \toprule
            Configuration & Throughput (tokens/s) \\
            \midrule
            RCCL ($8\times$AMD)
            & $10{,}718 \pm 141$ \\
            NCCL ($8\times$NVIDIA)
            & $20{,}889 \pm 20$ \\
            HetCCL ($8\times$AMD + $8\times$NVIDIA)
            & $30{,}640 \pm 229$ \\
            \bottomrule
        \end{tabular}%
    }
\end{table}

Table~\ref{tab:training_variability} reports three independent
runs for the LLaMA-1B configuration with ZeRO-3 and FP16
from Figure~\ref{fig:dl_perf}.
We discard the first 20 warm-up steps and average throughput
over the steady-state window in each run.
Across both tables, the coefficient of variation is at most
1.31\%. The results show small run-to-run variation and preserve
the performance trends reported in
Sections~\ref{eval-comm-perf} and~\ref{eval-dl-training}.


\section{Ablation study}
\label{sec:ablation-study}

In this section, we provide an ablation study, including the effect of the RDMA mechanism (Appendix~\ref{subsec:ablation-rdma}), and the effect of the GPU-aware load balancing technique used in \sys (Appendix~\ref{subsec:ablation-load-balancing}).

\subsection{Effect of RDMA}
\label{subsec:ablation-rdma} 

\begin{figure*}[ht] 
    \centering
    \includegraphics[width=0.65\linewidth]{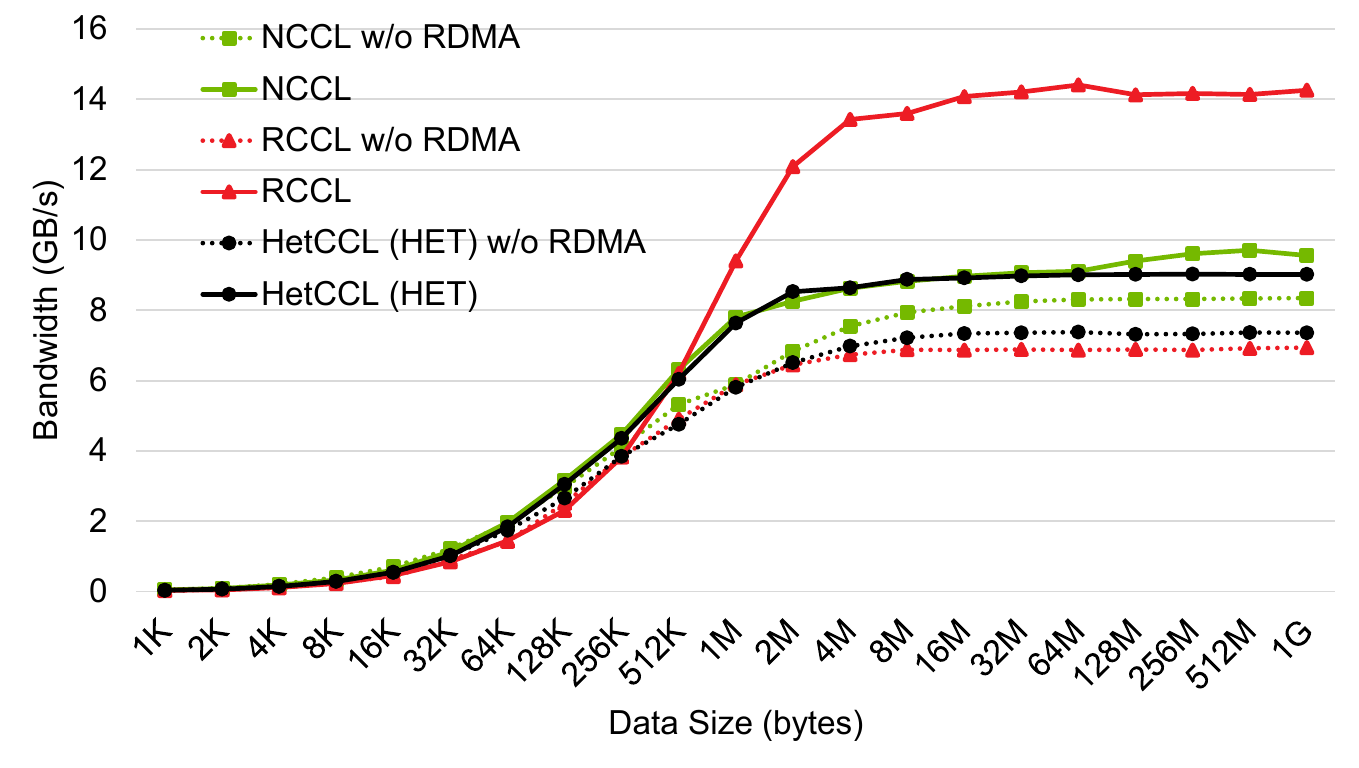}
    \caption{Performance impact of RDMA on point-to-point GPU communication.}
    \label{fig:rdma-ablation}
\end{figure*}

We include an ablation that isolates the contribution of the GPUDirect RDMA mechanism~\cite{nvidia_gpudirect, amd_directgma} on point-to-point communication performance. The non-RDMA baseline relies on host-staged communication, where data is explicitly copied from GPU memory to CPU memory before being transferred by the NIC, and then copied back to the destination GPU. This results in a data path of \texttt{GPU} $\rightarrow$ \texttt{CPU} $\rightarrow$ \texttt{NIC} $\rightarrow$ \texttt{CPU} $\rightarrow$ \texttt{GPU}.
In contrast, GPUDirect RDMA enables direct GPU-to-NIC and NIC-to-GPU transfers, resulting in a path of \texttt{GPU} $\rightarrow$ \texttt{NIC} $\rightarrow$ \texttt{GPU}, eliminating host involvement and achieving higher bandwidth, as shown in Figure~\ref{fig:rdma-ablation}.

\subsection{Effect of GPU-Aware Workload Balancing}
\label{subsec:ablation-load-balancing}

We present an ablation study on the GPU-aware workload balancing technique implemented in \sys for efficient training with heterogeneous multi-vendor GPUs. \sys assigns a larger micro-batch size to faster GPUs to mitigate load imbalance across devices. Specifically, NVIDIA GPUs are assigned approximately twice the micro-batch size of AMD GPUs, reflecting the approximately 2$\times$ higher profiled training throughput of NVIDIA V100 GPUs compared to AMD W7800 GPUs in our cluster. Although the theoretical peak FP16 performance is 112~TFLOPS for the NVIDIA V100 and 90.5~TFLOPS for the AMD W7800, we observe substantially lower effective hardware utilization on the AMD GPUs during end-to-end LLM training. This disparity is primarily attributed to AMD's relatively low maturity of the software stack and kernel-level optimizations, a phenomenon consistent with recent performance analysis of AMD's software ecosystem~\cite{ambati2025amd}.

Consequently, \sys prioritizes empirical profiling over naive theoretical metrics to ensure robust workload distribution in heterogeneous environments. The profiling run executes a small number of warm-up training iterations using the same model configuration and parallelization setup as the main training run, and measures the achieved throughput in tokens per second.

\begin{table}[ht]
\centering
\caption{Performance impact of load balancing on LLM training.}
\vspace{0.7ex}
\label{tab:ablation-load-balancing}
\resizebox{0.53\linewidth}{!}{
    \begin{tabular}{lcc}
    \toprule
    \textbf{Model} & \textbf{End-to-end speedup} & \textbf{Profiling overhead (s)} \\
    \midrule
    GPT 125M   & 1.22$\times$ & 1.4 \\
    GPT 355M   & 1.19$\times$ & 2.9 \\
    LLaMA 1B   & 1.09$\times$ & 26.6 \\
    LLaMA 3B   & 1.08$\times$ & 80.2 \\
    \bottomrule
    \end{tabular}
}
\end{table}

The experimental results in Table~\ref{tab:ablation-load-balancing} compare uniform micro-batch assignment with \sys's GPU-aware workload balancing, where faster GPUs are assigned more work.
Given the profiling-derived micro-batch assignment ratio of approximately 1:2, an ideal speedup of 1.5$\times$ might be expected.
However, the observed throughput improvements are consistently lower than this upper bound.
This gap primarily stems from the communication fraction in end-to-end training, which limits the extent to which compute-side load balancing can improve overall throughput. 

\begin{table}[ht]
\centering
\caption{End-to-end training latency breakdown across models.}
\vspace{0.7ex}
\label{tab:latency_breakdown}
\resizebox{0.7\linewidth}{!}{
    \begin{tabular}{lccc}
        \toprule
        \textbf{Model} & \textbf{Communication (\%)} & \textbf{Computation (\%)} & \textbf{Synchronization (\%)} \\
        \midrule
        GPT 125M & 13.26 & 86.72 & 0.02 \\
        GPT 355M & 16.96 & 83.03 & 0.01 \\
        LLaMA 1B & 22.85 & 77.14 & 0.01 \\
        LLaMA 3B & 29.09 & 70.91 & 0.00 \\
        \bottomrule
    \end{tabular}
}
\end{table}

To better understand this behavior, we break down the end-to-end training latency into communication, computation, and explicit synchronization. Since communication and computation can overlap in practice, we measure the cumulative duration of communication kernels and synchronization primitives and derive the computation portion by subtracting both from the total end-to-end training time. As shown in Table~\ref{tab:latency_breakdown}, while computation remains the dominant component, communication already constitutes a substantial fraction of the training step and becomes increasingly important as model size grows.
In particular, the communication portion increases from 13.3\% for GPT-125M to 29.1\% for LLaMA-3B.
This indicates that end-to-end throughput is increasingly constrained by communication rather than pure computation, which reduces the achievable computational performance gain from GPU-aware workload balancing.
The effect is especially pronounced for larger models, where larger gradients and heavier collective communication lead to diminishing speedups.

All experiments in this section use ZeRO stage 3, which represents the most communication-intensive configuration and therefore serves as a conservative evaluation setting. Although the main experiments were evaluated under data parallelism, this rank-level micro-batch adjustment mechanism is orthogonal to the parallelism strategy and can be applied to other forms of parallelism.

Table~\ref{tab:ablation-load-balancing} also reports the overhead of a short profiling job in \sys required to determine appropriate workload ratios. Even for the largest model (LLaMA-3B), the profiling cost is trivial (less than two minutes) and is amortized over long training runs consisting of thousands of training steps. We intentionally adopt a lightweight workload balancing strategy to minimize system complexity and overhead. While more sophisticated balancing approaches~\cite{guo2024cephalo} are possible, the results show that our profiling-based load balancing approach is sufficient to eliminate major straggler effects in practice.

\section{Impact Statement}
\label{sec:impact-statement}

By enabling communication across heterogeneous multi-vendor GPUs, our work aims to improve resource utilization and reduce the cost of DL model training. We believe this can lower the barrier to training LLMs in diverse computing environments. We do not identify any direct negative societal impacts specific to this work beyond the general risks associated with enabling more efficient training.

\newpage
\section*{NeurIPS Paper Checklist}

\begin{enumerate}

\item {\bf Claims}
    \item[] Question: Do the main claims made in the abstract and introduction accurately reflect the paper's contributions and scope?
    \item[] Answer: \answerYes{}
    \item[] Justification: The abstract and introduction clearly state the key contributions, proposed mechanisms, and main experimental findings.

\item {\bf Limitations}
    \item[] Question: Does the paper discuss the limitations of the work performed by the authors?
    \item[] Answer: \answerYes{}
    \item[] Justification: We discuss the limitations of this work and future directions in Section~\ref{sec:limitations}.

\item {\bf Theory assumptions and proofs}
    \item[] Question: For each theoretical result, does the paper provide the full set of assumptions and a complete (and correct) proof?
    \item[] Answer: \answerNA{}
    \item[] Justification: The paper does not include theoretical results.

    \item {\bf Experimental result reproducibility}
    \item[] Question: Does the paper fully disclose all the information needed to reproduce the main experimental results of the paper to the extent that it affects the main claims and/or conclusions of the paper (regardless of whether the code and data are provided or not)?
    \item[] Answer: \answerYes{}
    \item[] Justification: We describe the system configuration, experimental setup, implementation details, and evaluated workloads in Section~\ref{sec:eval} and Appendix~\ref{sec:experimental-details}.

\item {\bf Open access to data and code}
    \item[] Question: Does the paper provide open access to the data and code, with sufficient instructions to faithfully reproduce the main experimental results, as described in supplemental material?
    \item[] Answer: \answerYes{}
    \item[] Justification: We use public models and datasets. We provide the code with detailed README files in the supplementary material and plan to open-source it upon acceptance.

\item {\bf Experimental setting/details}
    \item[] Question: Does the paper specify all the training and test details (e.g., data splits, hyperparameters, how they were chosen, type of optimizer) necessary to understand the results?
    \item[] Answer: \answerYes{}
    \item[] Justification: We provide the experimental setup, model configurations, software stack, and parallelization details in Section~\ref{sec:eval} and Appendix~\ref{sec:experimental-details}.

\item {\bf Experiment statistical significance}
    \item[] Question: Does the paper report error bars suitably and correctly defined or other appropriate information about the statistical significance of the experiments?
    \item[] Answer: \answerYes{}
    \item[] Justification: Appendix~\ref{subsec:statistical_variability} reports the mean and standard deviation across three independent job launches for collective communication benchmarks and a representative end-to-end training configuration, together with the measurement protocol.

\item {\bf Experiments compute resources}
    \item[] Question: For each experiment, does the paper provide sufficient information on the computer resources (type of compute workers, memory, time of execution) needed to reproduce the experiments?
    \item[] Answer: \answerYes{}
    \item[] Justification: We report the hardware and software resources used for the experiments in Section~\ref{sec:eval}, Appendix~\ref{sec:experimental-details}, and supplementary material.
    
\item {\bf Code of ethics}
    \item[] Question: Does the research conducted in the paper conform, in every respect, with the NeurIPS Code of Ethics \url{https://neurips.cc/public/EthicsGuidelines}?
    \item[] Answer: \answerYes{}
    \item[] Justification: We fully follow the NeurIPS Code of Ethics.

\item {\bf Broader impacts}
    \item[] Question: Does the paper discuss both potential positive societal impacts and negative societal impacts of the work performed?
    \item[] Answer: \answerYes{}
    \item[] Justification: See Appendix~\ref{sec:impact-statement}.
    
\item {\bf Safeguards}
    \item[] Question: Does the paper describe safeguards that have been put in place for responsible release of data or models that have a high risk for misuse (e.g., pre-trained language models, image generators, or scraped datasets)?
    \item[] Answer: \answerNA{}
    \item[] Justification: The paper poses no such risks.

\item {\bf Licenses for existing assets}
    \item[] Question: Are the creators or original owners of assets (e.g., code, data, models), used in the paper, properly credited and are the license and terms of use explicitly mentioned and properly respected?
    \item[] Answer: \answerYes{}
    \item[] Justification: We properly cite the existing software, datasets, and models used in the paper, and follow their licenses and terms of use.

\item {\bf New assets}
    \item[] Question: Are new assets introduced in the paper well documented and is the documentation provided alongside the assets?
    \item[] Answer: \answerYes{}
    \item[] Justification: We provide the code with detailed README files in the supplementary material and plan to open-source it upon publication.

\item {\bf Crowdsourcing and research with human subjects}
    \item[] Question: For crowdsourcing experiments and research with human subjects, does the paper include the full text of instructions given to participants and screenshots, if applicable, as well as details about compensation (if any)? 
    \item[] Answer: \answerNA{}
    \item[] Justification: The paper does not involve crowdsourcing or research with human subjects.

\item {\bf Institutional review board (IRB) approvals or equivalent for research with human subjects}
    \item[] Question: Does the paper describe potential risks incurred by study participants, whether such risks were disclosed to the subjects, and whether Institutional Review Board (IRB) approvals (or an equivalent approval/review based on the requirements of your country or institution) were obtained?
    \item[] Answer: \answerNA{}
    \item[] Justification: The paper does not involve crowdsourcing or research with human subjects.

\item {\bf Declaration of LLM usage}
    \item[] Question: Does the paper describe the usage of LLMs if it is an important, original, or non-standard component of the core methods in this research? Note that if the LLM is used only for writing, editing, or formatting purposes and does \emph{not} impact the core methodology, scientific rigor, or originality of the research, declaration is not required.
    \item[] Answer: \answerNA{}
    \item[] Justification: The core method development in this research does not involve LLMs as any important, original, or non-standard components.

\end{enumerate}

\end{document}